\documentclass[pre,a4paper,showpacs,superscriptaddress,floatfix,twocolumn]{revtex4}

\usepackage{amsmath,epsfig,amsfonts,amssymb}

\newcommand{\rme}{ {\rm e} }
\newcommand{\rmd}{{\rm d} } 
\newcommand{\aref}[1]{\ref{#1}(a)}
\newcommand{\bref}[1]{\ref{#1}(b)}

\begin{document}
\title{Spreading in narrow channels}
\author{C. Dotti}
\author{A. Gambassi}
\author{M. N. Popescu}
\author{S. Dietrich}
\affiliation{Max-Planck-Institut f\"ur Metallforschung, Heisenbergstr. 3,
D-70569 Stuttgart, Germany}
\affiliation{Institut f\"ur Theoretische und Angewandte Physik,
Universit\"at Stuttgart, Pfaffenwaldring 57, D-70569 Stuttgart,
Germany}

\begin{abstract}

We study a lattice model for the spreading of fluid films, which are a few
molecular layers thick, in narrow channels with inert lateral walls.
We focus on systems connected to two particle
reservoirs at different chemical potentials, considering an attractive
substrate potential at the bottom, confining side 
walls, and hard-core repulsive fluid-fluid interactions.
Using kinetic Monte Carlo simulations we find a diffusive behavior.
The corresponding diffusion coefficient depends on the density and is bounded
from below by
 the free one-dimensional diffusion coefficient, valid for an inert bottom
 wall. These numerical results are rationalized   within the
 corresponding continuum limit.   
\end{abstract}

\pacs{02.50.-r, 05.70.Ln, 68.15.+e, 81.15.Aa}

\maketitle

\section{Introduction}
In recent years, substantial progress has been made in the development of
the "lab on a chip" concept, i.e., the 
integration of  many physical and chemical processes (e.g., transport through
micro-channels, mixing of different fluids, chemical reactions) into a
single device; entire laboratory setups, like a gas chromatograph, have been
miniaturized 
on a single chip (for a review see, e.g., Ref.~\cite{Giord_Cheng}). 
In this context, microfluidics is becoming a standard tool in
many applications, 
ranging from biology (see, e.g., Ref.~\cite{SeCaMa}) to the handling of toxic or
rare substances. Further scaling down 
to nanofluidics is expected to take place in the future~\cite{Mukho}.
Already now it is possible to sculpture channels with lateral dimensions of few tens
of nanometers \cite{Pears_Cumm} (for a review on such fabrication processes see
Ref.~\cite{Rev_Fabr})
and carbon nanotubes have been proposed as possible pipes in
nanofluidics~\cite{Supple_Quirke_03,Hum_Ras_Now_01}. 
Chemically patterned substrates have also been suggested as
a solution for directed transport, gating, mixing, or separation of fluids at
the micro- and nano-scale~\cite{Dietr_Pop_Rau_05}. In this case
the channel consists of a strip of 
wettable material embedded in a non-wettable substrate so that the fluid
flows along the wettable region and is laterally confined by the chemical
contrast.   

If one of the dimensions of a fluid film is comparable to the 
size of the fluid molecules, a hydrodynamical description of the film is no
longer justified~\cite{MD_early,Gleb_PRE,V_DCon}. In this case the discrete
nature of the fluid becomes important and the fluid cannot be treated as a
continuum in the confined direction. 
In order to investigate such systems one possible approach is to carry out
computer simulation 
of discrete models,  e.g., molecular dynamics, kinetic Monte Carlo (KMC), or
lattice Boltzmann simulations; 
recent work in this direction includes fluids in carbon nanotubes 
\cite{Supple_Quirke_03} and on chemically patterned substrates 
\cite{Kop_Lo_Dietr_Rausc}.

With the scaling down of microfluidic devices one has to deal with and may
exploit the ultrathin precursor film which spreads ahead of the bulk fluid.
Experimental studies have shown that
in some cases such precursor films have molecular
thickness~\cite{Exp1,Exp2,Exp3,Exp4,Exp5,Exp6,Exp7,Exp8}. 
The spreading of such monolayers has been
studied using
a two-dimensional lattice gas Ising model~\cite{Gleb_PRE,Burlatsky,Gleb_eal}
in which a half-space is occupied by a particle reservoir. 
Recent KMC simulations and a continuum analysis~\cite{Pop_Dietr_04} of
that model 
provided results in good qualitative agreement with available experimental data,
and a further extension to the case of
chemically patterned substrate has been proposed~\cite{Pop_Dietr_Osh_05}. 

Fluids in narrow channels have been investigated theoretically in the context of
 single-file diffusion, i.e., when fluid particles cannot overtake each other (see, e.g.,
 Refs.~\cite{Chou1,Chou2,Chou3,Chou4,MonPer,BhaJeNich,DemStSu,MarTe,Bech,troppiA}).
 Such systems show the interesting feature of non-diffusive 
 behavior of tracer particles, which stimulated 
 experimental (see, e.g., Refs.~\cite{Bech,troppiA}) and
 numerical~\cite{Chou3,MonPer,BhaJeNich,DemStSu} interest.  
Here we present a lattice model for ultrathin films in which 
multiple occupancy of a site is allowed (generalizing the single-occupancy
model of Refs.~\cite{Gleb_PRE,Burlatsky,Gleb_eal}) and in which the 
substrate-particle attractive interaction is decaying as a power law, whereas the
particle-particle interaction is assumed to be hard-core repulsive only. This
mimics the case in which the
fluid-substrate interaction strongly dominates over the actual attractive long-range part of the
fluid-fluid interaction.  
Based on the phenomena occurring in this minimalistic model, the extension to
the case in which the attractive part of the fluid-fluid interaction  
is relevant will be presented elsewhere. 
We shall restrict our analysis to a
one-dimensional model, which can effectively describe fluids in extremely narrow
channels with a width which is less than twice the particle diameter. The
sidewalls act to confine the particles. The corrugation of the substrate
potential both at the bottom and at the sides is incorporated effectively by
considering a lattice model for the particles. Due to the small thickness of the
channel the transversal variation of the substrate potential can be
ignored. This model is supposed to mimic not only molecular fluids but also
colloidal particles in solution, with the colloidal particle setting the
length scale.
 
We 
discuss both the initial dynamics, in which a fluid film fed by a reservoir 
gradually fills the channel, and the steady state, in which  the fluid film
in the channel is in contact with two reservoirs at different chemical
potentials.

The paper is organized such that in Sec.~\ref{Mod} we define the
model whereas in Sec.~\ref{MC_sim} the results of our Monte Carlo simulations are
presented. The analyses of the diffusion-like dynamics and of the steady-state
properties are presented in Sec.~\ref{furt_an}. In 
Sec.~\ref{CL} we discuss the mean-field continuum limit of the model and
rationalize, within this approximation, the results for the diffusion
coefficient presented in Sec.~\ref{furt_an}.   
Section~\ref{Conc} summarizes the main findings and provides our
conclusions.  

\section{ The model}\label{Mod}
Before specifying the rules defining the model, we further describe
the general physical picture of the type of systems we have in mind. 

As stated above, the fluid is assumed to be confined to a narrow, effectively one-dimensional
channel. The sidewalls are very high compared with the fluid
particle diameter, so that the fluid cannot spill out of
the channel.  The channel walls act on the particles such that only the
vertical variation of the  substrate potential matters. The left and the right
end of the channel are connected to a feeding and absorbing particle reservoir,
respectively, and the channel is initially empty. 
The fluid film inside the channel is taken to be compact, i.e.,
molecules are densely packed to form vertical columns without vacancies. This corresponds to the
case in which the substrate is strongly attractive and 
vacancies inside columns are eliminated on a time scale much shorter than the typical time for
exchanges of particles between columns. We describe these exchanges
in terms of
\emph{rates}, which are related to the change of the energy of the system due
to the corresponding move. Particle exchanges 
between columns and particle insertions and removals near the reservoirs
are the only processes we consider. We assume that neither evaporation nor condensation 
takes place inside the channel. 
This minimalistic model aims at identifying general aspects and the main
qualitative features of 
fluids spreading in a strongly confined geometry, rather than providing an accurate
description of a particular physical situation. 

Inspired by this picture, in Subsec.~\ref{Conf_H} we
specify the configuration space and the corresponding energy
function. 
In Subsec.~\ref{rates} the dynamical rules (i.e., the allowed changes of the
configurations and the associated rates) governing the time evolution in the bulk are
discussed, whereas Subsec.~\ref{BC} deals with the definition of the
dynamics at the feeding and absorbing boundaries.

\subsection{Configurations and Hamiltonian}\label{Conf_H}
\begin{figure}[hc]
\centering
\epsfig{figure=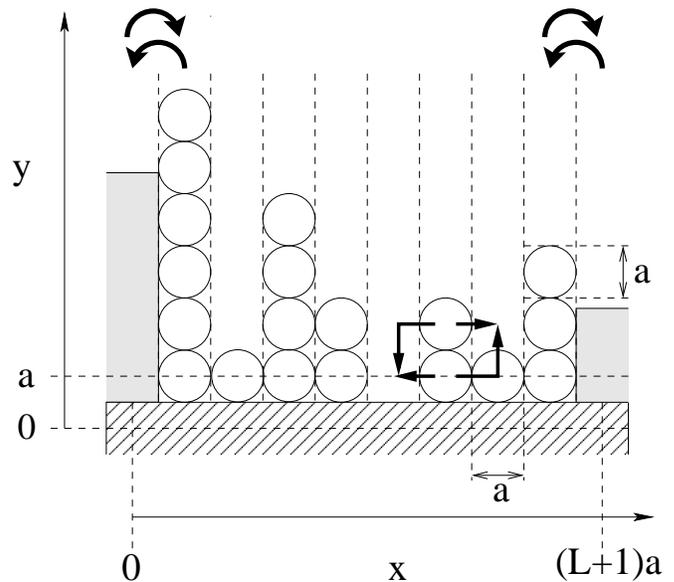,width=\columnwidth}
\caption{A typical configuration of the model. The possible moves
  in the bulk are indicated by straight
  arrows, while the curved arrows denote reservoirs-system exchanges.
 The substrate, including the exclusion zone of its top layer,
  corresponds to the hatched area. The grey areas at $x \leq 0$ and $x \geq
  (L+1)a$ indicate reservoirs and the fluid particles are shown as circles. }\label{Config}
\end{figure}

The model is defined  
on a one-dimensional ($D=1$) lattice, with sites $[0,\ldots,(L+1)a]$. 
The distance $a$ between two consecutive sites is
assumed to be equal to the effective diameter of a fluid particle, which is set by the
hard-core repulsion between the particles. In the following, 
the site indices and the distances will be expressed in units of $a$. 
The two sites indexed with $0$ and $L+1$ are the
boundaries of the left and right particle reservoirs, respectively. The other
sites, $[1,\dots,L]$ (called `bulk' in the following), represent the channel of
length $L$ (assumed to be long, i.e., $L \gg 1$). 

 At every site $x \in [0,L+1]$ an occupation number 
$n_x \in \mathbb{N}_0$ specifies the number of particles piled in
the column $x$ (see Fig.~\ref{Config}). Within the
column, particles centers are located at integer $y$ positions. 
Accounting for fluid-substrate hard-core repulsion we consider
the position $y=0$  
as passing through the centers of the particles forming the top layer of the 
substrate. If the diameter of the substrate particles differs from that of the
fluid particles, one may introduce an extra parameter to characterize the
position of the fluid-substrate contact layer; for simplicity, however, we assume that the
fluid particles in the first layer are located at $y=1$ (see Fig.~\ref{Config}).

The substrate is assumed to be uniform, and, consistent with
 our one-dimensional model, two-dimensional
semi-infinite in the $y<0$ - direction. We denote the (attractive) pair interaction
between a substrate particle and a fluid particle by $U_{sf}^{(p)}$, resembling
dispersion forces:
\begin{equation}\label{LJ-s}
U_{sf}^{(p)} (d) = \left\{ \begin{array}{ll} -\frac{w_{sf}}{d^6}, & \textrm{for} \: d \geq 1, \\
    \infty, & \textrm{for} \: d<1, \end{array} \right.
\end{equation}
where $d$ is the dimensionless distance in units of $a$ between the substrate particle located at
$(x',-y' \leq 0)$, and the fluid particle located at $(1 \leq x \leq L,y \geq 1)$.
In the case of pairwise additive interactions, for a semi-infinite substrate ($x' \in
\mathbb{R} $, $y' \in \mathbb{R}_+$) and in the continuum limit ($d \gg 1$),
this leads to a total substrate potential 
\begin{equation}\label{Tot_sp}
U_{sf} (y) = - w_{sf} \int_{0}^{\infty} \rmd y' \int_{-\infty}^{\infty} \rmd x' \frac{1}
{\left[ (y'+y)^2+(x-x')^2 \right]^{3}} ,
\end{equation}
i.e.,
\begin{equation}\label{Usp}
U_{sf} (y) = \left\{ \begin{array}{ll} -\frac{w'_{sp}}{y^4}, &  \textrm{for}
    \: y \geq 1 \: ,  \\
    \infty, & \textrm{for} \: y < 1 \: , \end{array} \right.
\end{equation}
where $w'_{sf} = \frac{3 \, \pi}{32} \, w_{sf}$. 
 Within this ansatz, the particle-substrate interaction in Eq.~(\ref{Usp})
 depends on the height $y$ of the particle only. The energy of the fluid configuration $\{
n_1,\dots,n_L \}$ exposed to the substrate potential $U_{sf}$ is thus given by
\begin{equation}\label{hsp}
H_{sf}=\sum_{x=1}^{L} \sum_{y_x=1}^{n_x} U_{sf}(y_x),
\end{equation}
where the inner sum is defined to be $0$ if $n_x=0$. 
Note that, following the discussion at the beginning of the present section,
we assume that columns are always densely packed, so that configurations are
as depicted in Fig.~\ref{Config}: since the configurations are 
characterized completely by a succession of numbers $\{n_1,\dots,n_L\}$, the energy is
a function of these numbers only, as in Eq.~(\ref{hsp}).

The same form [Eq.~(\ref{LJ-s})] of the pair potential is assumed for
the fluid particle - fluid particle interaction, where the corresponding interaction
strength is denoted by $w_{ff}$. 
Each pair of particles separated by a distance $d_{ff} \geq 1$ contributes to
the particle-particle energy, so that the total energy due to particle-particle
interactions can be written as 
 \begin{equation}\label{hpp}
\begin{split}
 & H_{ff}=\frac{1}{2} \sum_{x=1}^{L} \sum_{x'=1}^{L} \sum_{y_x=1}^{n_x} \sum_{y_{x'}=1}^{n_{x'}} \\
 & \hfill U_{ff}^{(p)}\left(d_{ff}=\sqrt{(x-x')^2+(y_x-y_{x'})^2} \right), 
\end{split}
 \end{equation} 
with $U_{ff}^{(p)}(0)=0$ and the sums over
$y_x$ and $y_{x'}$ are taken to be zero if $n_x=0$ or $n_{x'}=0$.
The total energy function is $H[C]=H_{ff}+H_{sf}$ where
$C\equiv\{n_1,\dots,n_L\}$ characterizes completely each configuration. Note that this part of the
Hamiltonian 
is restricted to the bulk; in general the reservoir-bulk interactions should also be
accounted for separately. 
In the special case of the absence of long-range particle-particle interaction, i.e., $w_{ff}=0$,
both the bulk and the reservoir-bulk contributions vanish, and the energy is
$H[C]=H_{sf}$. As mentioned in the Introduction, in the following
we shall discuss only this situation; the case $w_{ff} \neq 0$ will be
presented elsewhere.  
   
\subsection{The rates and the dynamics}\label{rates} 

In this subsection we define and discuss the rates which govern the stochastic
dynamics in the bulk, i.e., for $x \in [1,L]$.
The dynamics at the boundaries, $x=0$ and $x={L+1}$, will be discussed in
the following subsection.  

We assume that each particle in the column $x$ may jump into one of the nearest neighbor (NN) columns
$x+1$ or $x-1$. We introduce the rate $r_{CC'}(y,y')$,
which is the rate for a particle in column $x$ and at given height $y$, to jump
to the next column $x+1$ and at height $y'$. Within our aforementioned model
assumption this process involves an instantaneous column height reduction by
one in column $x$ and a column height increase in column $x+1$. This also
means that the jumping particle is considered to be able to squeeze into
column $x+1$ at position $y'$ by pushing the particles above this position up
by one unit; on the other hand if $y'$ is above the top particle of column
$x+1$, it falls down in order to form again a compact column. Accordingly the
configurations $C,C'$,  
\begin{equation}\label{configurations}
\begin{split}
C &= \{ n_1,\dots,n_x,n_{x+1},\dots, n_L \} \\ 
C' &= \{ n_1,\dots,n_{x}-1,n_{x+1}+1,\dots, n_L \},
\end{split}
\end{equation}
represent the initial
and the final configurations, for any pair $y$,$y'$. Analogous considerations
can be carried out for moves from $x+1$ to $x$, where the above
configurations are interchanged. Therefore, within our model,
the rates $r_{CC'}(y,y')$ depend \emph{only} on the initial and final configurations
$C$ and $C'$, respectively. Accordingly, the dependence on $y$,$y'$ is dropped.
We introduce the dimensionless rate $\tilde{u}_{CC'}$, which is also assumed
to depend only on $C$,$C'$,
\begin{equation}\label{nodim_rate}
\tilde{u}_{CC'} = \frac{r_{CC'}}{\nu_0}, 
\end{equation} 
where $\nu_0$ fixes the time-scale of the model and we assume it to be independent
of the source and target columns filling, i.e., the same for \emph{any}
particle in the source column.
($\nu_0$ can be interpreted as the
rate for a particle to jump to the NN column if the energy happens to be
unchanged by the move.) In the following, times are measured in units of
$\nu_0$, i.e., the dimensionless simulation time $t$ corresponds to an actual time $t_a =
t/\nu_0$. 
 
We choose the rates $\tilde{u}$ for \emph{all} possible moves from column $x$ to column
$x+1$ such that detailed balance
\begin{equation}\label{detailed}
\frac{\tilde{u}_{CC'}}{\tilde{u}_{C'C}} = \rme^{-\beta \Delta H[C,C']}   
\end{equation}
is obeyed, 
where $\Delta H[C,C']=H[C']-H[C]$ is the energy difference between the final ($C'$) and
the initial ($C$) configuration. Detailed balance has been chosen in order to
ensure that thermal equilibrium is reached in
the long-time limit, if the two reservoirs of particles at the right and the left of
the channel are set to the same chemical potential. 
A possible choice that satisfies the detailed balance condition is 
\begin{equation}\label{rate}
\tilde{u}_{CC'} =  \rme^{-\frac{\beta}{2} \Delta H[C,C']}. 
\end{equation}
The chosen form of the rates [Eq.~(\ref{rate})] includes
 both ``slow'' ($\Delta H >0$) and ``fast'' ($\Delta H <0$) processes, and we
 implicitly assume that it captures essential features of the real dynamics.

The rate $\tilde{u}_{CC'}$ is the same for any particle in the source column $x$, so
that the \emph{total} rate $u_{CC'}$ for a
column to decrease its occupation number by one, while a given NN column 
increases its own occupation number by one, is 
\begin{equation}\label{rate2}
u_{CC'}  = n_x \tilde{u}_{CC'} = n_x \rme^{-\frac{\beta}{2} \Delta H[C,C']}.
\end{equation}
The rates
in Eq.~(\ref{rate2}) are defined on the space of configurations
specified by occupation numbers only. Detailed balance still holds and the
corresponding Boltzmann statistical weight is
\begin{equation}\label{mod_bw}
p_B({n_1,\dots,n_x,\dots,n_L}) \propto \, \frac{\rme^{-\beta H}}{n_1!\dots n_L!}
\end{equation}
which accounts for "particle undistinctness" by dividing the Boltzmann factor by
$n_1!,\dots, n_L!$ where $n_x!$ is the number of choices to label the $n_x$
particles in each $x \in [1,\dots, L]$ column.  
In the case of the particle-substrate
interaction described by the Hamiltonian in Eq.~(\ref{hsp}), the rate in
Eq.~(\ref{rate2}) has the following explicit form:
\begin{equation}\label{rate3}
u\left(n_x,n_{x+1}\right) = n_x
\exp{\left\{\frac{\beta}{2} \left[ \frac{w'_{sp}}{(n_{x+1}+1)^4}
    - \frac{w'_{sp}}{n_x^4} \right] \right\} }.
\end{equation}
This formula emphasizes that
$u$ depends only on the occupation numbers of the initial and the target column. The
notation is such that the first argument stands for the source column (here
located at $x$ with occupation $n_x$), while
the second argument represents the target column (here at $x+1$ with occupation $n_{x+1}$). 

Assuming that the dynamics leads to a diffusion-like behavior (as will be
discussed in Sec.~\ref{furt_an}), some 
qualitative features of the diffusion coefficient as a function of the local
density can be anticipated from the general properties of the rates in Eq.~(\ref{rate3}).
First, consider the situation in which both $n_x=n$ 
and $n_{x \pm 1}=m$ are large compared to $(\beta w'_{sp})^{1/4}$. Then the exponent in
Eq.~(\ref{rate3}) is very small and $u(n,m) \rightarrow n$, so that the
model reduces to free particles diffusing in $D=1$. 
The same conclusion holds for $n=m+1$, in which case the exponent
is zero leading to $u(n,m)=n$. In general,
 $u(n,m) \gtrless n$ if $n \gtrless m+1$; accordingly, jumps from high
 columns to low columns 
are faster than in the free case, while the opposite processes are
slower. This means that diffusion, which tends to smooth out density gradients, is 
\emph{enhanced} by the exponential factor in Eq.~(\ref{rate3}).
Since at low densities most of the configurations are composed
either of empty columns or of columns occupied by one particle, the most probable rate is
$w(1,0)=1$, which results in free diffusion at low densities.

These considerations lead to the conclusion that the diffusion
coefficient is expected to exhibit a peak
at relatively low densities, because the rates exhibit the maximal difference with
free diffusion rates if the target column is empty.

Before passing to the definition of the dynamics at the boundaries, we briefly comment on
similar models which have been considered in the literature. 
In Refs.~\cite{Coc-thiv1,Coc-thiv2} a class of dynamical models, to which our model belongs, is
introduced and studied. In this class of models the rates depend on both
the source and the target column, they do not necessarily 
satisfy detailed balance, and jumps occur not only between NN. (These models are
known in the literature as misanthropic processes.) The main result
of Refs.~\cite{Coc-thiv1,Coc-thiv2} is that under certain
conditions in the infinite square lattice it is possible to obtain an exact
expression for the steady-state distribution.
A concise summary of these 
results can be found in Refs.~\cite{Godr1,Godr2}, where their relevance for the
non-equilibrium dynamics of interacting particles has been stressed.  
Applied to our case, the
results in Refs.~\cite{Coc-thiv1,Coc-thiv2} recover the equilibrium Boltzmann distribution 
in Eq.~(\ref{mod_bw}), with the Hamiltonian defined in Eq.~(\ref{hsp}), but
do not provide information on the dynamics and steady-state distribution if
chemical drive, caused by different chemical potential for the two reservoirs
at the boundaries, is applied.

\subsection{Dynamics at the boundaries}\label{BC} 
We  consider now the dynamics at the boundaries and discuss two possible
implementations. The first choice is
to fix the occupation number of the columns $0$ and $L+1$
at values $n_0$ and $n_{L+1}$, respectively,
and to impose, with some additional assumptions, the same dynamics as in the
``bulk''. The boundary dynamics changes the
occupation number $n_1$ of the first
column of the system, according to Eq.~(\ref{rate3}), while the
occupation number $n_0$ is unchanged by the move, and the same holds at
$x=L$. One can physically motivate such a choice by assuming
that the particle exchanges within the reservoir are so fast, so that a
particle extracted from the reservoir is immediately replaced.
Under this assumption the density of particles in the reservoir is simply $n_0$.
While this choice seems to be rather natural, as
explained in the following the equilibrium (i.e., for $n_0=n_{L+1}$)
properties display an unexpected feature, i.e., 
a jump discontinuity in the density between the reservoirs and the system.

\begin{figure}
\centering
\epsfig{figure=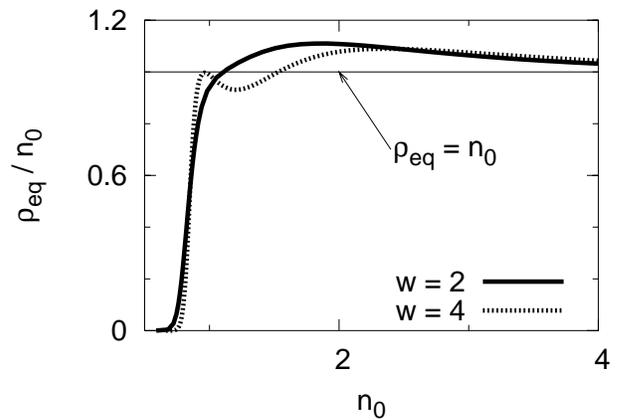,width=\columnwidth}
\caption{The equilibrium density $\rho_{eq}$
  [Eqs.~(\ref{GC_eq})-(\ref{res_dens}) and~(\ref{fug})], divided by the reservoir 
  occupation number $n_0$, as a function of $n_0$ for $w=2$ and
  $w=4$, respectively.}\label{eq_dens}
\end{figure}

The total Hamiltonian in Eq.~(\ref{hsp}) is a sum
of single-column terms, so that the equilibrium grand canonical distribution
factorizes:
\begin{equation}\label{GC_eq}
P_{eq}(\{ n_1,\dots,n_L \}) = \frac{1}{Z(w, \mu)} 
 \prod_{k=1}^{L} p_w(n_k) \, \rme^{-\mu n_k},
\end{equation}
where
\begin{equation}\label{Z-eq}
Z(w, \mu) = \left( \sum_{n=0}^{\infty}  p_w(n) \,\rme^{-\mu n} \right)^L
\end{equation}
is the total partition function, $\mu=\beta \tilde{\mu}$ is the dimensionless
chemical potential, with $\tilde{\mu}$ as the actual chemical potential, 
$p_w$ is a non-normalized single-column statistical weight,
\begin{equation}\label{p-eq}
p_w(m) = \frac{1}{m!} \exp{\left[ 2 w \, h(m) \right]},
\end{equation}
$w = \beta w'_{sp}/2$ is a dimensionless quantity, and 
\begin{equation}\label{HN}
\begin{split}
& h(m) = \left\{ \begin{array}{ll} {\displaystyle \sum_{k=1}^{m} \frac{1}{k^4}}, 
    & \textrm{for} \: n \geq 1, \\
    0 , & \textrm{for} \: n=0. \end{array} \right.
\end{split}
\end{equation}
The chemical potential $\mu$ controls the mean density of particles in the system,
\begin{equation}\label{res_dens}
\rho_{eq}(z) \equiv \frac{N_{eq}}{L} = \frac{z}{L} \partial_z \ln{\left[Z\left(w,z
    \right) \right]}=  \frac{{\displaystyle \sum_{n \geq 1} n\, p_w(n)
    z^n}}{{\displaystyle \sum_{n \geq 0} p(n) z^n}},  
\end{equation}
 where $N_{eq}$ is the mean total number of particles in the system, 
and $z=\rme^{-\mu}$ is the fugacity.
Detailed balance for the rates at the left reservoir reads
\begin{equation}\label{DB_res}
\begin{split}
& P_{eq}(\{ n_1,\dots,n_L \}) \, u(n_0,n_1) =\\
& = P_{eq}(\{ n_1+1,\dots,n_L \}) \,u(n_1+1,n_0)
\end{split}
\end{equation}
where the probability per unit time of inserting a particle into the column at
$x=1$ is compared with the corresponding
probability per unit time of removing the particle in the same column. A
similar condition has to hold at the right end $x=L$ of the system.
Combining Eq.~(\ref{DB_res}) with Eqs.~(\ref{rate3}) and~(\ref{GC_eq}) leads to 
\begin{equation}\label{fug}
z = n_0
\exp{\left\{-w \left[\frac{1}{n_0^4}
    +\frac{1}{(n_0+1)^4} \right]\right\} }.
\end{equation} 
Equations~(\ref{res_dens}) and~(\ref{fug}) give the equilibrium density $\rho_{eq}$
in the system as a function of $n_0$. As expected, if $n_0$ 
is large $\rho_{eq}$ coincides with $n_0$: in Eq.~(\ref{fug})
$n_0 \gg w^{1/4}$ implies $z \approx n_0$, and in Eq.~(\ref{res_dens}) $n_0
\gg w^{1/3}$ implies 
$p_w(n) \approx \exp{(2 w \zeta(4) )}/n!$, so that for $n_0 \gg
\max{(w^{1/3},w^{1/4})}$ Eq.~(\ref{res_dens}) reduces to
\begin{equation}
\rho_{eq} \approx \rme^{-z} z\, \frac{\partial}{\partial z} \rme^z = z \approx
n_0, \quad  n_0 \gg \max{(w^{1/3},w^{1/4})}.
\end{equation} 
In the range of substrate potential strength we investigated ($0.5<w<5$) the approximation
$\rho_{eq} \approx n_0$ is valid if $n_0>5$. As the substrate potential
strength is increased, this threshold increases, while it tends to zero for $w \rightarrow 0$.
For densities lower than the threshold, the
reservoir occupation number $n_0$  does not 
coincide with the density $\rho_{eq}$ of the equilibrium system as obtained from
Eqs.~(\ref{GC_eq})-(\ref{res_dens}) and~(\ref{fug}) (see Fig.~\ref{eq_dens}).
For example, in the case $n_0=3$ one has
$\rho_{eq} \simeq 3.19$ for $w=2$ and $\rho_{eq} \simeq 3.23$ for $w=4$. 
These densities are in very good agreement with the simulation data for the
density $\rho_1$ in the first column (see, c.f., Fig.~\bref{St_Profs}).
In the simulations we investigated a non-equilibrium situation in
which the two reservoirs at the boundaries have different occupation numbers
($n_0 \neq n_{L+1}$); 
nevertheless we recover the equilibrium density in the first column. 
This shows how it is possible to control the densities at the first ($x=1$)
and last ($x=L$) site by varying the occupation numbers $n_0$ and
$n_{L+1}$. In order to obtain
arbitrary densities, it is necessary to take $n_0$
and $n_L$ to be continuous, thus loosing the direct physical interpretation
of these parameters. Moreover, as shown in Fig.~\ref{eq_dens}, $\rho_{eq}$
drops sharply for $n_0 \lesssim 0.8$, and in the range $ 0 \lesssim
\rho_{eq} \lesssim 0.8$ a high 
numerical accuracy would be required to determine the corresponding value of
$n_0$.  

These two problems can be solved by generalizing the dynamics at the
boundaries as follows. In Eq.~(\ref{rate3}) the terms depending on  
$n_0$ and $n_{L+1}$, i.e., the properties of the reservoirs, are replaced by
constants $\alpha$, $\gamma$, $\delta$, and $\kappa$ in the following way:
\begin{equation}\label{Grates}
\begin{split}
u_\alpha(n_1) & = \alpha \exp{\left[\frac{w}{(n_1+1)^4}\right]}, \: 
 u_\gamma(n_1) = \gamma \, n_1 \exp{\left[-\frac{w}{n_1^4}\right]},  \\
u_\delta(n_L) &= \delta \exp{\left[\frac{w}{(n_L+1)^4}\right]}, \:  
u_\kappa(n_L) = \kappa \, n_L \exp{\left[-\frac{w}{n_L^4}\right]},  
\end{split}
\end{equation}
where $u_\alpha$ is the rate for particle insertion into column $n_1$ from the
left reservoir,
$u_\gamma$ is the rate for particle removal from column $n_1$ into the
left reservoir and $u_\delta, u_\kappa$ are the corresponding rates at $x=L$. 
Imposing detailed balance [Eq.~(\ref{DB_res})] for the rates defined in
Eq.~(\ref{Grates}) gives
\begin{equation}\label{Gen_res}
\rme^{-\mu} = z = \frac{\alpha}{\gamma} = \frac{\delta}{\kappa}.
\end{equation}
In the non-equilibrium case the fugacity of the right reservoir,
denoted as $z_{L+1}=\delta/\kappa$, and the one of the left reservoir, i.e.,
$z_0=\alpha/\gamma$, are different ($z_0 \neq z_{L+1}$). 
Using these two fugacities in Eq.~(\ref{res_dens}), the densities of the two
corresponding equilibrium systems are found; we \emph{define} them to be the
reservoir densities. In simulations we proceed backwards:
first we choose $\rho_0=\rho_{eq}(z_0)$ and $\rho_{L+1}=\rho_{eq}(z_{L+1})$,
and then we find the 
corresponding ratios by inverting Eq.~(\ref{res_dens}). Setting
$\gamma=\kappa=1$ implies $\alpha=z_0$, $\delta=z_{L+1}$ so that the
inversion is simpler. 

\section{ Monte-Carlo simulations}\label{MC_sim}

The \emph{continuous time} dynamics defined by the rules described in
Subsecs.~\ref{rates} and~\ref{BC} 
is simulated using a Kinetic Monte Carlo (KMC) method~\cite{Binder}.
At every step an increment $\Delta t$ for the time variable is drawn from the
distribution
\begin{equation}
\begin{split}
& P(\Delta t)  = \frac{1}{S\left(n_0, \dots, n_{L+1} \right)} \exp{\left[  S\left(n_0, \dots, n_{L+1} \right)
    \Delta t \right]} ,
\\
& S\left(n_0, \dots, n_{L+1} \right)  = \sum_{x=0}^{L} \left[ u(n_x,n_{x+1}) +
    u(n_{x+1},n_{x}) \right],
\end{split}
\end{equation}
where $S$ is the total rate to leave the configuration $\{n_0,\dots,n_{L+1}\}$.
The move to perform is then chosen according to the weight $u/S$ of its rate. We used a
classical N-fold way 
algorithm~\cite{N-fold}, which has the advantage that the selected moves are
accepted without rejection. 
The model depends on four parameters: the
substrate interaction strength $w=w'_{sp} \,
\beta/2$ in units of the thermal energy, the two boundary densities $n_0$ and
$n_{L+1}$, and the length $L$ of the 
system. The simulations have been performed up to a maximum time $\tau_{tot}$
and quantities have been measured after the initial time $\tau_0$.
The simulations covered both the spreading and the steady-state regime
and in both cases we sampled the same set of quantities. In order to keep the notation
simple we indicate averages always with $\langle \cdot \rangle$, but, as  described in
the following, the meaning of the symbol is different in the two situations considered.   
 
The mean density of particles, i.e., the density at site $x$ and time $t$ is defined as 
\begin{equation}\label{density}
\rho(x,t) = \langle n_x(t) \rangle,
\end{equation}
while the total mean number of particles (or mean total mass) is
\begin{equation}\label{M}
M(t) = \sum_{x=1}^{L}\langle n_x(t) \rangle.
\end{equation}
The transport properties have been studied using the integrated particle
current at site $x$ and time $t$,  
\begin{equation}
J(x,t,\Delta t) =
\Delta n_{x,x+1}(t,t+\Delta t) - \Delta n_{x+1,x}(t,t+\Delta t),  
\end{equation}
where $\Delta n_{x,x'}(t,t+\Delta t)$  is the number of particles jumping  
 from site $x$ to site $x'$ within the time interval $\Delta t$. We define a
 mean instantaneous current as
\begin{equation}\label{defcurr} 
\langle j(x,t) \rangle = \lim_{\Delta t \rightarrow 0} \frac{\langle J(x,t,\Delta t)
\rangle}{\Delta t}.
\end{equation}

\subsection{Spreading}
 
\begin{figure*}

\begin{minipage}{0.5\textwidth}
\epsfig{figure=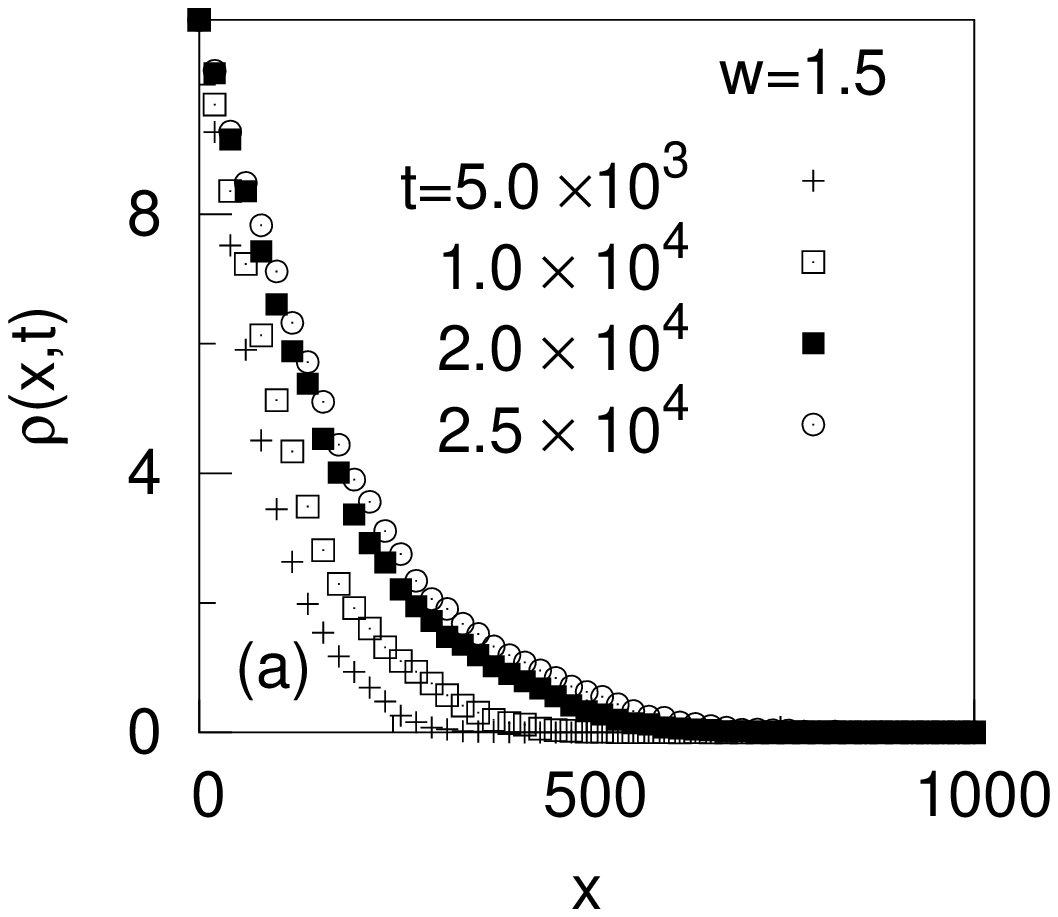,width=0.8\textwidth}
\end{minipage}\begin{minipage}{0.5\textwidth}
\epsfig{figure=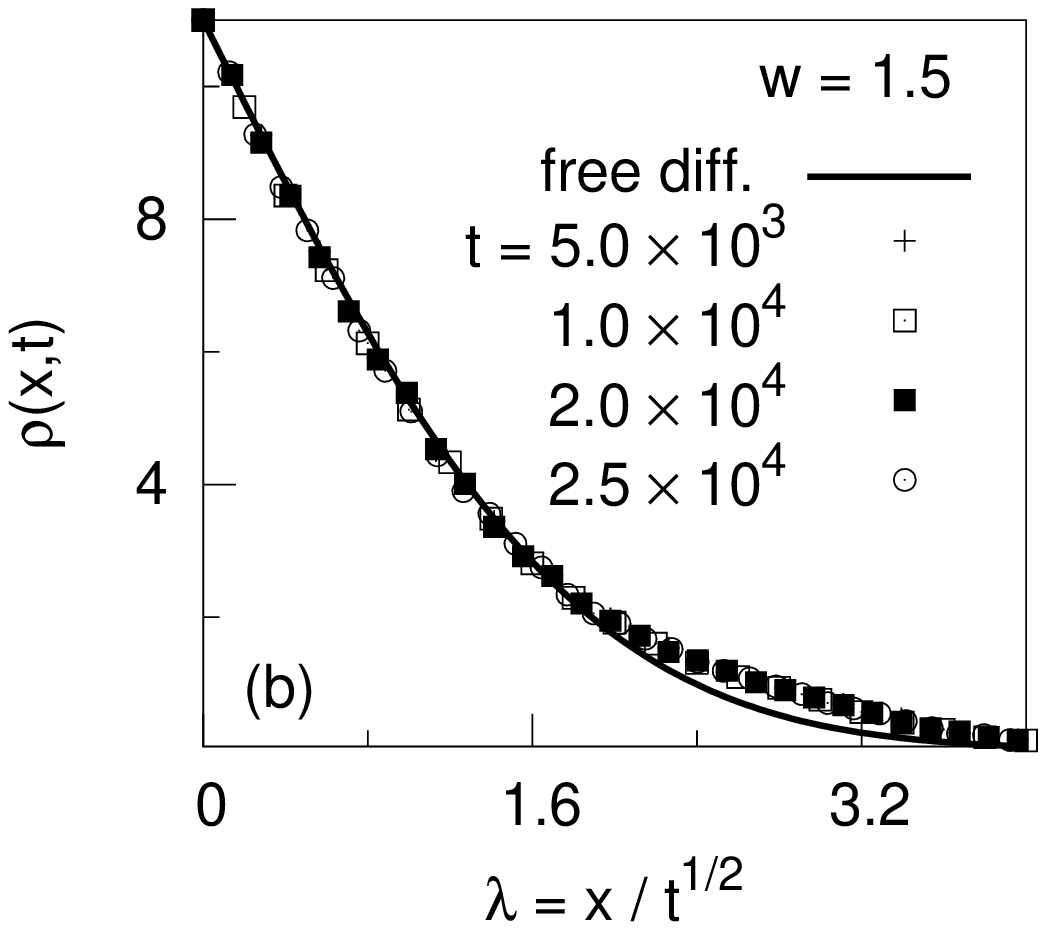,width=0.75\textwidth}
\end{minipage}

\caption{Time-dependent density profiles for spreading
   in a system with $L=1000$, for $n_0=11$, $n_{L+1}=0$, and $w=1.5$ at times
   $t=5 \times 10^3$ 
   ($+$), $10^4$ ($\boxdot$), $2\times 10^4$ ($\blacksquare$), and $2.5 \times 10^4$
   ($\odot$) as a function of $x$~(a), and  of $\lambda=x/
   \sqrt{t}$~(b), respectively. The solid line indicates the scaling function
   for free diffusion (i.e., $w=0$). }\label{Mv_Profs}
\end{figure*}
 
\begin{figure}
\centering
\epsfig{figure=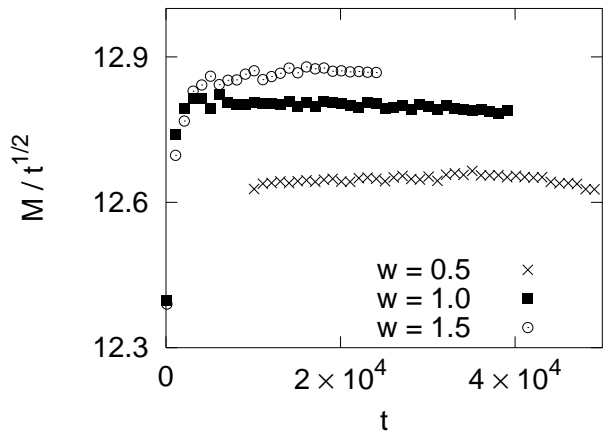,width=\columnwidth} 
\caption{Total mass $M/\sqrt{t}$ for spreading as a function of time $t$ for different values of the
  interaction $w$ with the substrate: $w=0.5$ (total simulation time $\tau_{tot}=5 \times 10^4$ and 
  initial sampling time $\tau_0=10^4$) ($\times$),
  $w=1.00$ ($\tau_{tot}=4 \times 10^4$, $\tau_0=0$) ($\blacksquare$),  and
  $w=1.50$ ($\tau_{tot}=2.5 \times 10^4$, $\tau_0=0$) ($\odot$) For all symbols
  $L=1000$, $n_0=11$, and $n_{L+1}=0$. }\label{mass}
\end{figure}

For the spreading regime the right reservoir has been converted into a particle
sink by setting $n_{L+1}=0$, while $n_0=11$ and $L=1000$. In the
simulations performed with these values of the parameters the 
leakage of particles through the sink is negligible for times $t \lesssim 
10^4$. We studied both the initial-time dynamics by setting $\tau_0=0$ and
$\tau_{tot} \leq 10^4$ and the long-time behavior 
in which both reservoirs play a role  (see, c.f., Fig~\ref{mass}, $w=0.5$); in this
latter case we have chosen $\tau_{tot}=5 \times 10^4$ and $\tau_0=10^4$ in
order to reduce the CPU and memory requirements. 
In the spreading regime we implemented the simulation average by
drawing different sequences of (pseudo-) random numbers while keeping  the
initial condition fixed so that in this case $\langle \cdot \rangle $ is the ensemble
average; the typical number of runs we averaged over is $2000$. 
We studied the shape of the density profile $\rho(x,t)$, defined in
Eq.~(\ref{density}), as a function of the interaction strength $w$ in the
range $0.5 \lesssim w \lesssim 1.5$.

Since for $w=0$ the dynamics reduces to free
diffusion, it is natural to check if in the general case (i.e., $w\neq 0$) the
profiles show a diffusive scaling. 
Plotting them as a function of $ \lambda =x / \sqrt{t}$ we indeed obtain a
collapse of data measured at different times, as shown in Fig.~\ref{Mv_Profs}.  
The rescaled profiles are similar to
the one for free diffusion, except for a small bend
at densities around $1$ (see Fig~\bref{Mv_Profs}), which depends on the
interaction strength $w$.  
The diffusive scaling is confirmed by the time evolution of
the total mass [Eq.~\eqref{M}] shown in Fig.~\ref{mass}, which is
expected to evolve as $M \propto \sqrt{t}$. We observe
deviations from this behavior only at short
times, when the boundary dynamics dominates, and at long times, when the
leakage of particles through the sink becomes relevant.

\subsection{Steady state}

\begin{figure*}
\begin{minipage}{0.5\textwidth}
\epsfig{figure=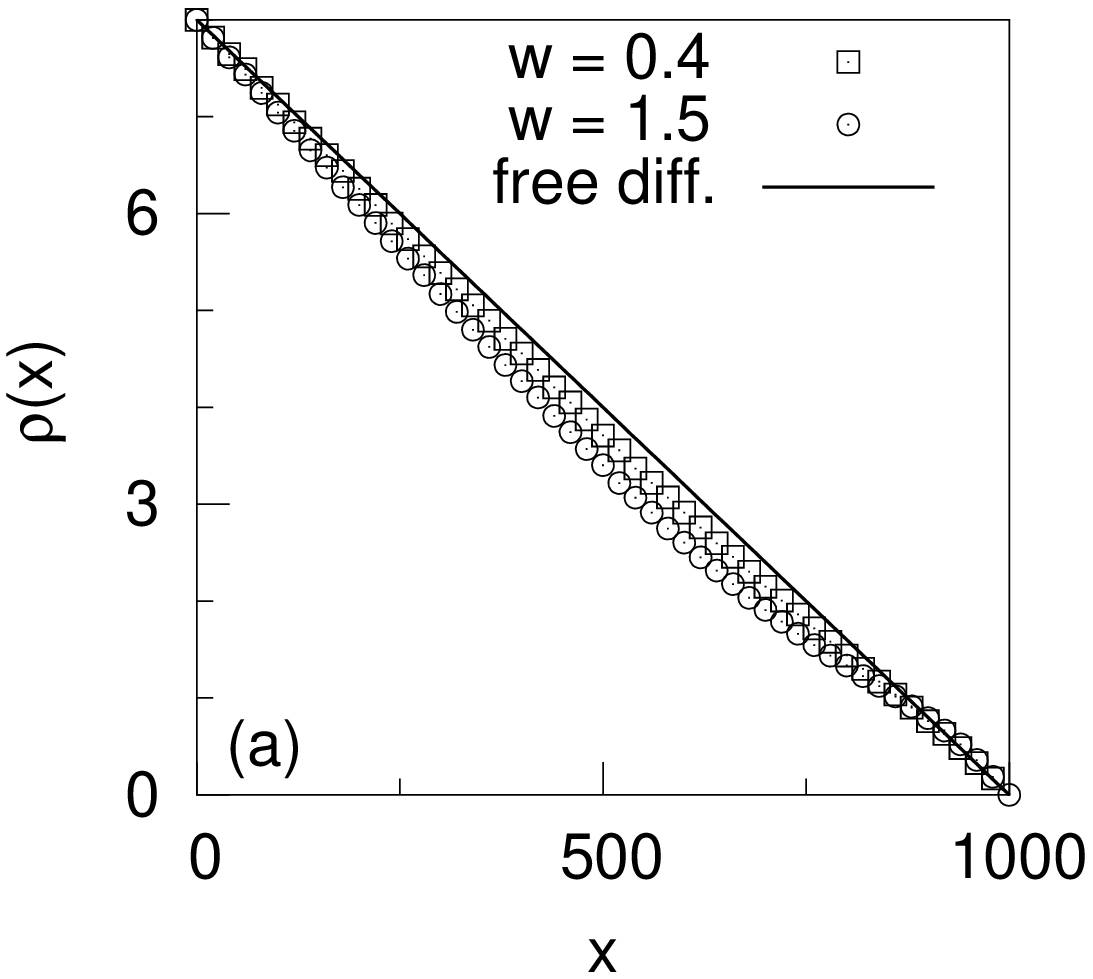,width=0.83\textwidth} 
\end{minipage}\begin{minipage}{0.5\textwidth}
\epsfig{figure=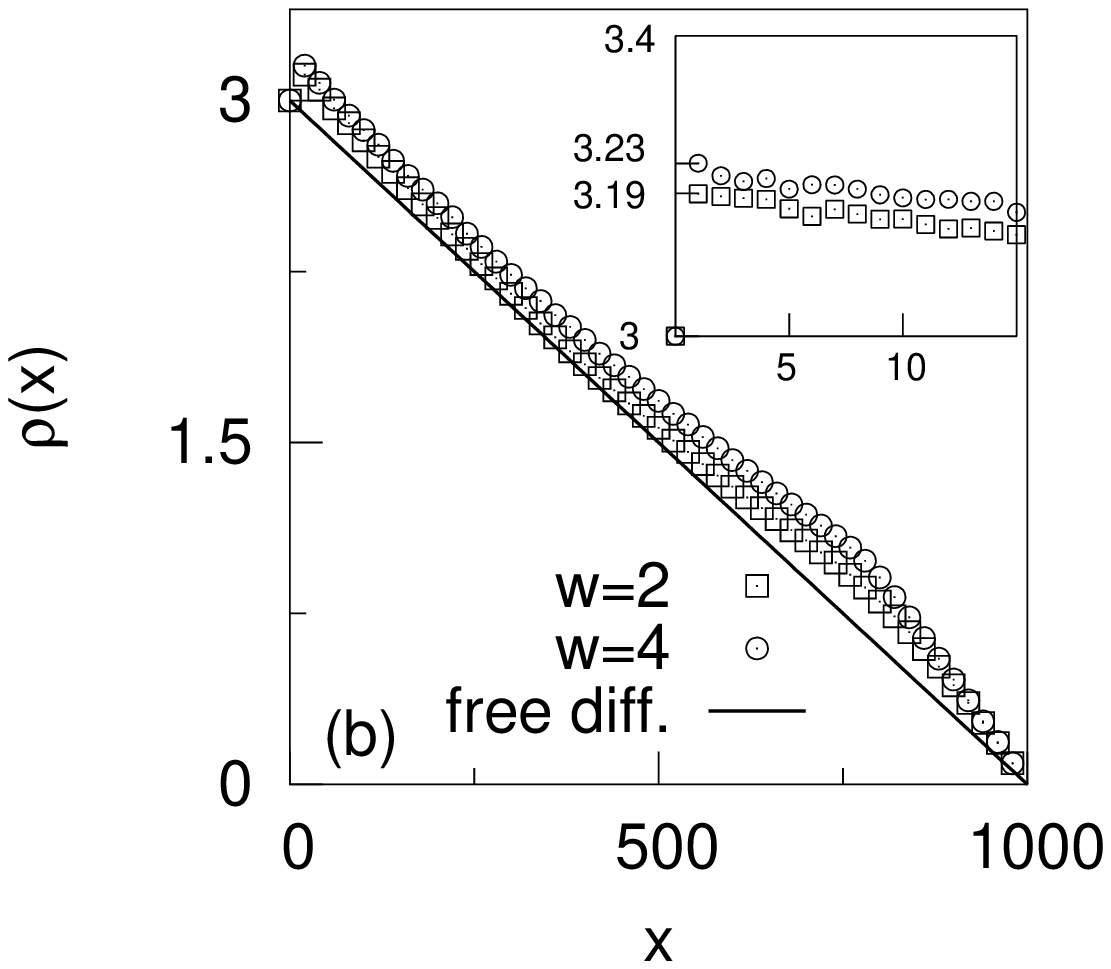,width=0.87\textwidth}
\end{minipage}

 \caption{  Steady-state density profiles in a system of
   length $L=1000$ for (a)~$w=0.4$ ($\boxdot$) and $w=1.5$
   ($\odot$), $n_0=8$, $n_{L+1}=0$; (b)~$w=2$
   ($\boxdot$) and $w=4$ ($\odot$), $n_0=3$, $n_{L+1}=0$. The
   inset in~(b) is a close-up view of the vicinity of the left
   reservoir showing $\rho_{eq} \simeq \rho_1 >n_0$
   [$\rho_{eq} \simeq 3.19$ for $w=2$ and $\rho_{eq} \simeq 3.23$ for $w=4$;
   see Eqs.~(\ref{res_dens}) and~(\ref{fug}) and the discussion in the main 
   text]. In both~(a) and~(b) the free diffusion is  
   indicated by a full line. }\label{St_Profs}

 \end{figure*}
The steady state is reached after running the simulation for an initial thermalization
time $\tau_0$ ($\tau_0 \approx 10^5$ for $L=1000$, chosen by checking that for $t>\tau_0$
the observables are time independent) and 
saving the configurations generated every  sampling
time interval $\tau_s$, with $\tau_s = 200$ for $L=1000$. The  average
$\langle \cdot \rangle$  for the observables defined above is taken over this
set of configurations. The choice for $\tau_s$ is a
compromise between speed and having as small correlations between
the $N_s$ measurements as possible.  We
assume that the total simulation time $\tau_{tot} = N_s \tau_s + \tau_0$ is sufficient to explore a 
significant part of the phase space, so that
the performed average coincides with the average over the (unknown) steady
state distribution. Note that this assumption is justified because 
no signs of dynamical phase transitions (which would introduce extremely long time scales) are
found in the simulations.

 The density $\rho(x)$ [Eq.~(\ref{density})] exhibits a
 profile smoothly interpolating between the two reservoirs
 (see~Fig.~\ref{St_Profs}) and slightly deviating from the
 corresponding free diffusion profile which is a straight line. 
This deviation is considered in more
 detail in Appendix~\ref{App2}, where its dependence on both the interaction
 and the reservoirs densities is analyzed. 
In the steady state both reservoirs play a role and finite-size
 effects have to be checked; it turns out that for $L>200$ there is no
 detectable dependence of 
 the data on the particular value of $L$ other than a trivial rescaling of
 the density profile.    

We have also determined the current $\langle j \rangle $ defined in
Eq.~(\ref{defcurr}): in the steady state $\langle j \rangle$
does not depend on $t$, so that $J(x,t,\Delta t)  =
\langle j \rangle \, \Delta t$ for any sufficiently large time interval $\Delta t$. $J$
can be obtained by measuring the flux of particles 
between any two sites $x$ and $x+1$, because in the steady state the current
$\langle j \rangle$ does not depend on $x$ due to local particle conservation. The 
instantaneous steady state current $\langle j(x) \rangle$ (a dependence on
$x$ is indicated to recall the random fluctuations around the mean value
$\langle j \rangle$) can then be obtained from 
\begin{equation}\label{stj}
\langle j(x) \rangle= \frac{J(x,0,\tau_{tot})-J(x,0,\tau_0)}{N_s \tau_s}.
\end{equation}
 Note that in Eq.~(\ref{stj})
 the  current integrated over the thermalization time, $J(x,0,\tau_0)$, which 
 depends on $x$ and $t$, has been subtracted.
 We have calculated $\langle j(x) \rangle$ for $x \in [1,\dots,L]$ leading to $\langle j
 \rangle = L^{-1} \sum_{x=1}^{L} \langle j(x) \rangle $.        

\section{\label{furt_an} Analysis of the simulation data}

\subsection{\label{diff} Methods to determine the diffusion coefficient}
\begin{figure}

\end{figure}
Guided by the results of the KMC simulations of the microscopic model we
expect that
a continuum (in space and time) description for the behavior of the model at long times and large
spatial scales is possible, i.e., that the hydrodynamic limit exists and is well defined. 
A rigorous proof has been provided for a small number of models (see, e.g., Ref.~\cite{Jona}).
In the present case such an explicit derivation appears to be a difficult task.

Assuming that the particle density $\rho$ and the current $\langle j \rangle$,
 defined in Eqs.~(\ref{density}) and~(\ref{defcurr}), are smooth functions of
 the position $x$ and of the time $t$,
the local conservation of particle density in the bulk (which is implicit in
 the dynamics of the model) is expected to take the form of a continuity equation:
\begin{equation}\label{CE-cont}
\partial_t \rho(x,t) = - \partial_x \langle j(x,t) \rangle.
\end{equation}   
The results of the simulations strongly indicate a diffusive scaling at long times and large spatial
scales, i.e., $\rho (x,t) =
\bar{\rho}(x/\sqrt{t})$, suggesting that the dynamics amounts to non-linear diffusion:
\begin{equation}\label{NL-diff}
\begin{split}
& \langle j(x,t) \rangle = - D(\rho) \partial_x \rho(x,t) \: \Rightarrow \: \\
& \partial_t \rho(x,t) =  \partial_x \left[ D(\rho) \partial_x \rho(x,t) \right]. 
\end{split}
\end{equation}
Based on the density profile $\rho(x,t)$ from the simulations, it is possible
to extract the function $D(\rho)$ from the data in the
spreading and the steady state regime, respectively, as follows. 
\begin{itemize}
\item{\bf Spreading.}
Using the scaling behavior 
$\rho(x,t) = \bar{\rho} (\lambda=x/\sqrt{t})$ (see the results in
Sec.~\ref{MC_sim}), Eq.~(\ref{NL-diff}) reduces to
\begin{equation}\label{NL-scal}
 \lambda \frac{\rmd}{\rmd\lambda} \bar{\rho}(\lambda) = 
 \frac{\rmd}{\rmd \lambda}\left[ D(\bar{\rho}(\lambda)) \frac{\rmd}{\rmd \lambda} \bar{\rho}(\lambda) \right].
\end{equation}
Assuming that $\frac{\rmd \bar{\rho}}{\rmd \lambda} D(\bar{\rho(\lambda)})$
and $\bar{\rho}(\lambda)$ vanish for $\lambda \rightarrow \infty$, which is
supported by the simulation data ($L \gg 1$ is an approximation for $L \rightarrow \infty$),
integrating Eq.~(\ref{NL-scal}) and inverting $\bar{\rho}(\lambda)$ into
$\lambda(\bar{\rho})$, one finds
\begin{equation}\label{Dn}
D(\bar{\rho}) =   \frac{\rmd}{ \rmd \bar{\rho}} \lambda(\bar{\rho})
\int_{0}^{\bar{\rho}} \rmd \rho \, \lambda(\rho).  
\end{equation}
This method might be inaccurate for small densities due to a systematic effect.
For $x \approx L$ the profile bends in order to fulfill the condition
$n_{L+1}=0$, so that its derivative $\frac{\rmd}{\rmd \lambda} \bar{\rho}$
is larger than the derivative of a profile $\bar{\rho}_\infty$ in the infinite
lattice: $\frac{\rmd}{\rmd \lambda}  
\bar{\rho}_\infty <  \frac{\rmd}{\rmd \lambda} \bar{\rho}$.
The integral in Eq.~(\ref{Dn}) is also 
underestimated, since $\lambda(\rho \rightarrow 0) \nrightarrow
\infty$. These effects lead to an underestimated  diffusion coefficient for
small values of $\rho$. The most
severe effect is probably due to the derivative
$\frac{\rmd}{ \rmd \lambda} \bar{\rho}(\lambda)$, while the errors in
the integral can be partially corrected by using larger lattice sizes. 

\item{\bf Steady state.}  In the stationary state the current and the density
  are constant with respect to time, so that Eq.~(\ref{NL-diff}) leads to 
\begin{equation}\label{Dnum}
D(\rho)  = -\frac{\langle j \rangle}{\rho\:'},      
\end{equation}
where $\rho\:'(x) = \frac{\rmd}{\rmd x} \rho(x)$.
Note that the computation of $D$ in this regime does not require any assumptions
on $\rho$ and $\rho \: '$, as in the previous case, and therefore no systematic
deviations from the actual diffusion coefficient are expected. 
\item{\bf Steady state in quasi-equilibrium.}
A steady state deviating only slightly from the equilibrium is
realized by imposing reservoir densities which differ only slightly.
Equation~(\ref{Dnum}) leads to
\begin{equation}\label{intD}
\langle j \rangle=\frac{1}{L}   \int_{\rho_{L}}^{\rho_{1}} \rmd \rho \, D(\rho),
\end{equation}
where $ \rho_1$ and $\rho_{L}$ are the densities at the first and the last site,
respectively, and $L$ is the length of the system.
For $\left| \rho_{1}-\rho_{L} \right| \ll \rho_{1}$ one has
\begin{equation}\label{SS-D}
D\left(\frac{\rho_{1}+\rho_{L}}{2}\right) \approx 
\frac{L \, \langle j \rangle}{\left( \rho_{1}-\rho_{L} \right )}.  
\end{equation}

\end{itemize}
\subsection{\label{in_diff} Results from the spreading regime}
\begin{figure*}
\centering
\begin{minipage}{0.5\textwidth} 
\epsfig{figure=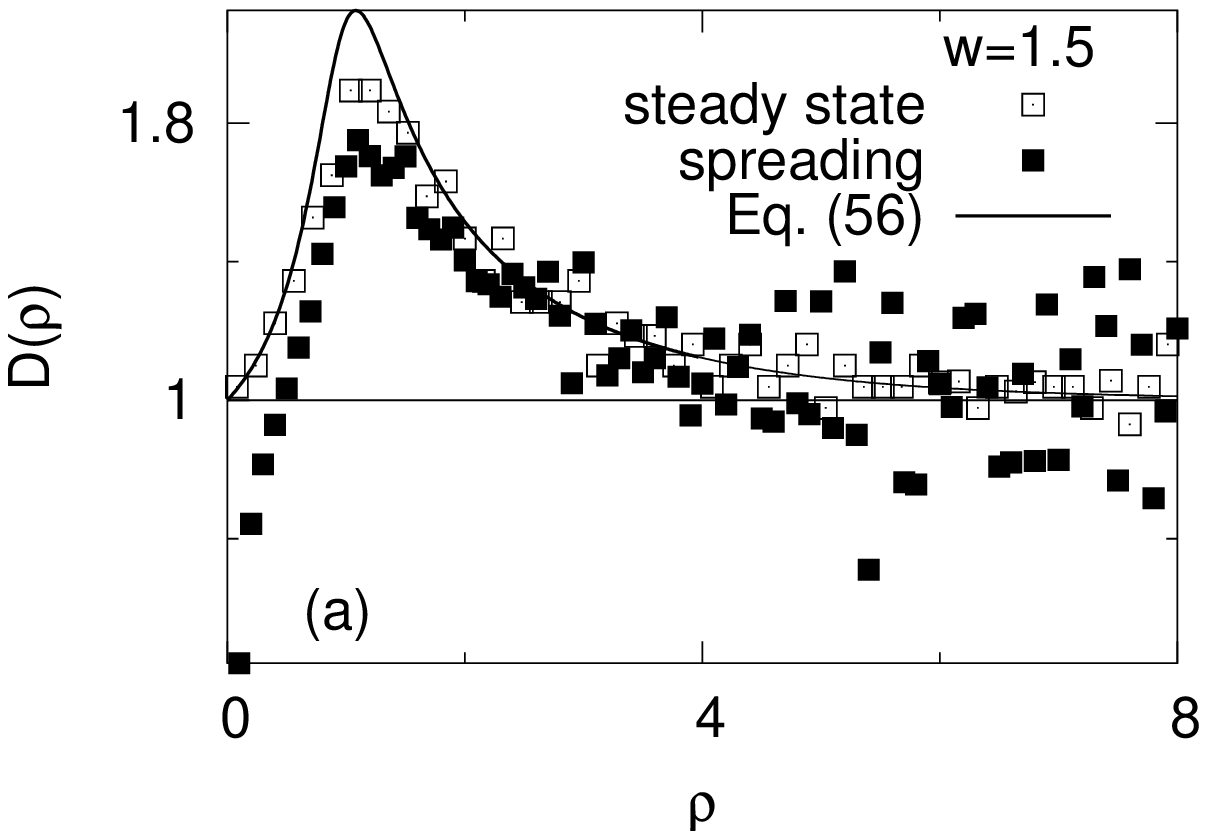,width=0.8\textwidth}\end{minipage}\begin{minipage}{0.5\textwidth}
\epsfig{figure=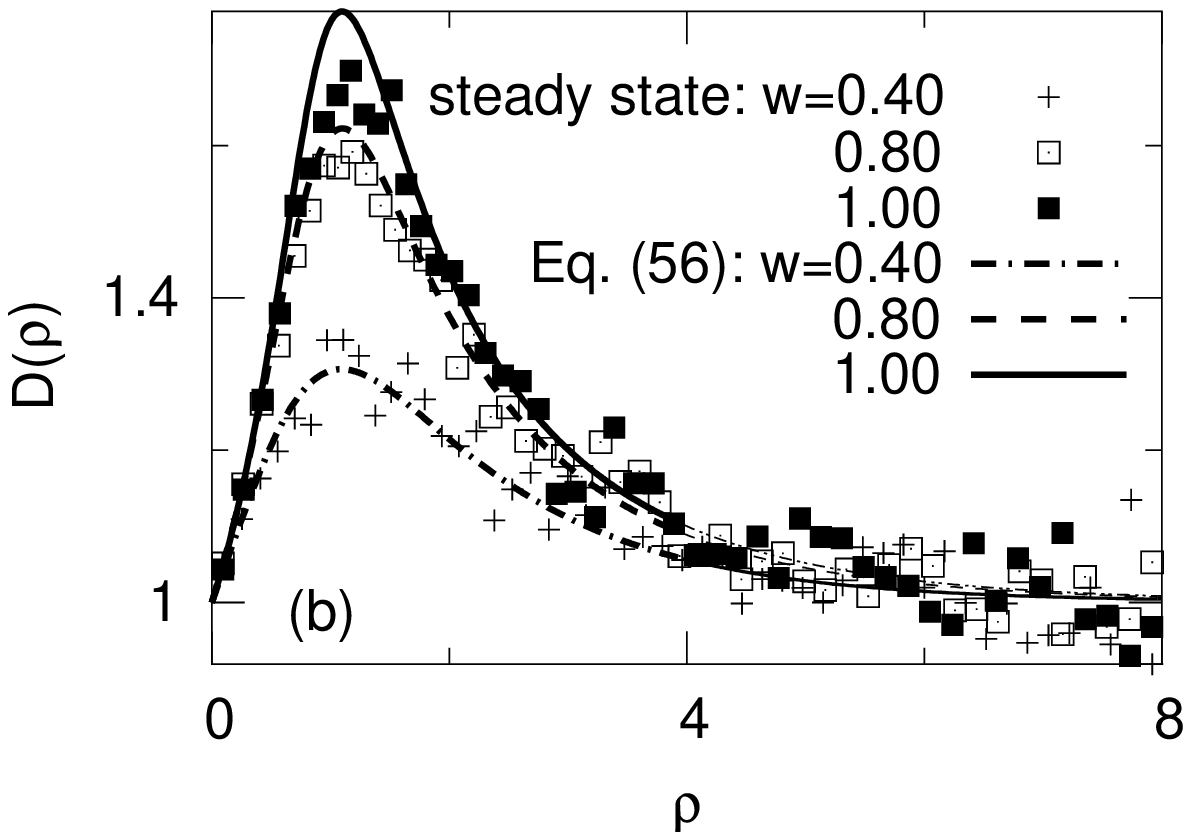,width=0.8\textwidth}
\end{minipage}
\caption{ (a)~Nonlinear diffusion coefficient $D$ as a function of the density,
  obtained from the simulation data in the spreading
regime ($\blacksquare$) and in the steady state ($\boxdot$), as well as the corresponding
analytical result [full line, Eq.~(\ref{Dmfin})]. All the data correspond to
$w=1.5$ and $L=1000$.\\
(b)~Nonlinear diffusion coefficient [Eq.~(\ref{Dnum})] from steady-state simulation
  data for $w=0.40$ ($+$), $w=0.80$ ($\boxdot$), and $w=1.00$ ($\blacksquare$) ($L=1000$).  Analytical
  results (lines) for $D(\rho)$ are calculated from Eq.~(\ref{Dmfin}).}\label{Diff_staz}
\end{figure*}
In order to extract $D(\rho)$ from the numerical data for given $w$ and $L$,
we have considered all the profiles in the scaling regime.
We have binned the $\rho$ axis with $\Delta \rho = 0.05$, averaged
all the values $\lambda$ in each bin, and evaluated the function $\lambda(\rho)$ 
by interpolation of the resulting data points, while $\frac{\rmd}{
  \rmd \lambda} \bar{\rho}(\lambda)$  has been obtained by finite differences.
The results for $D(\rho)$ obtained by using Eq.~(\ref{Dn}) are shown in Fig.~\aref{Diff_staz}. 
While it appears that, for large values of $\rho$,  $D(\rho) \rightarrow 1$ as expected from the
corresponding discussion in Subsec.~\ref{rates}, for
$\rho \rightarrow 0$ the diffusion coefficient goes to zero due to the systematic error in 
the derivative  $\frac{\rmd}{\rmd \lambda} \bar{\rho}(\lambda)$, as explained
in Subsec.~\ref{diff}. 
The noise at large values of $\rho$ is due to determining the derivatives
numerically, because for large $\rho$ the spatial fluctuations of the density are
stronger.    

The diffusion coefficient is peaked and the substrate potential enhances
diffusion (see Subsec.~\ref{rates}). The position of the peak
($\rho \simeq 1$) cannot be predicted by qualitative arguments, but is in the
range of low densities, as expected from the discussion in Subsec.~\ref{rates}.  

\subsection{\label{Steady} Results from steady-state and quasi-equilibrium regimes}
\begin{figure*}
\centering
\begin{minipage}{0.5\textwidth} 
\centering
\epsfig{figure=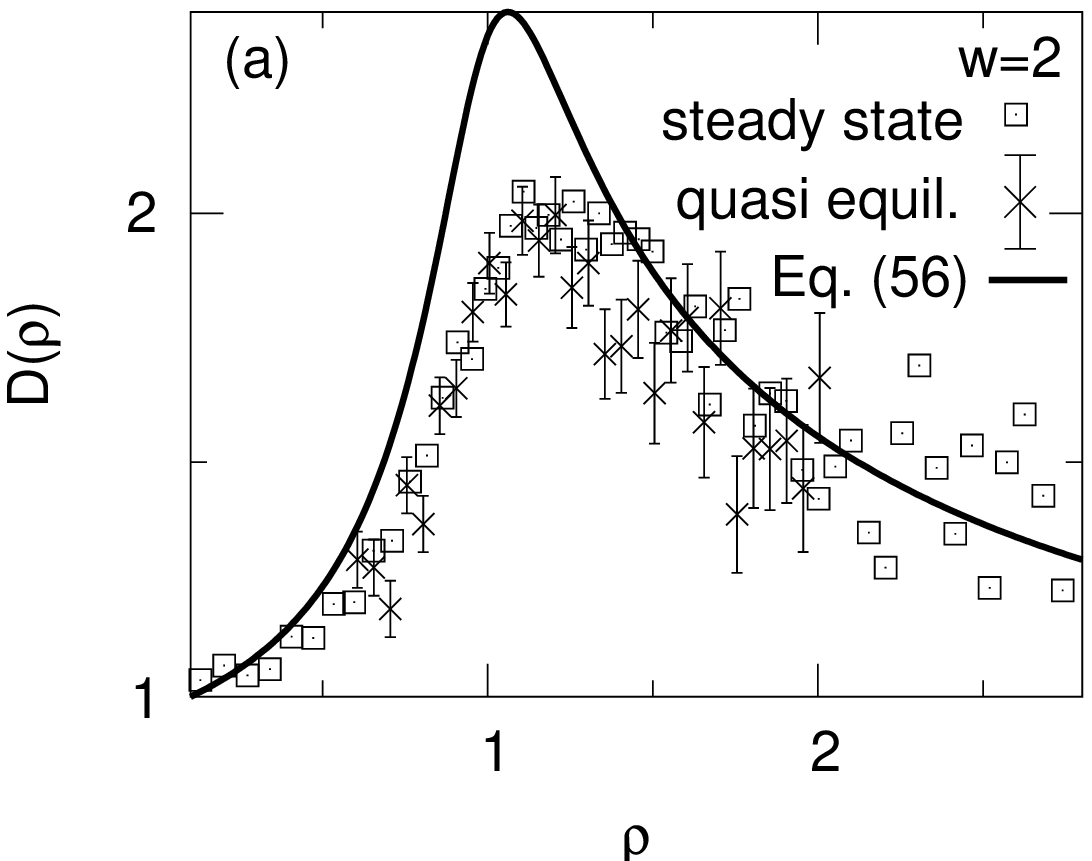,width=0.81\textwidth}
\end{minipage}\begin{minipage}{0.5\textwidth}
\centering 
\epsfig{figure=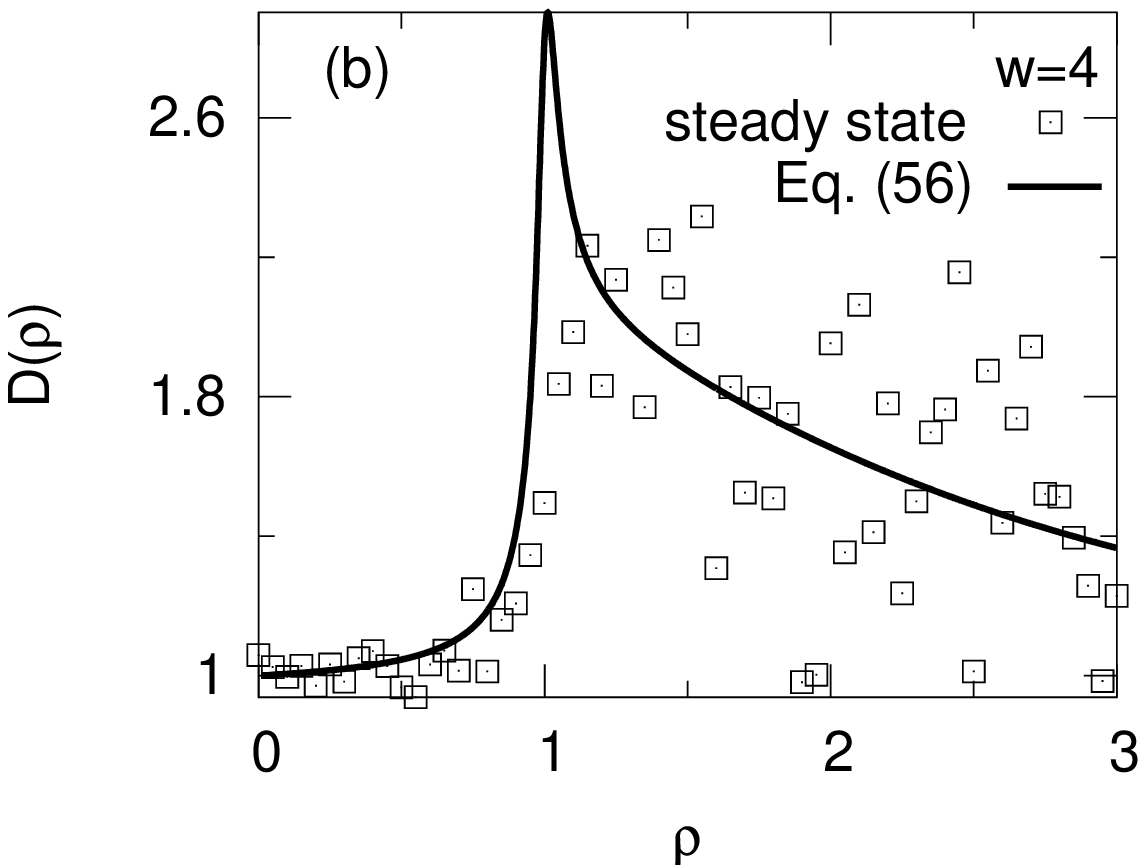,width=0.88\textwidth}
\end{minipage} 
\caption{Nonlinear diffusion coefficient $D(\rho)$ for large values of $w$.
  The open squares ($\boxdot$) are obtained from
  simulation data in the steady-state regime [Eq.~(\ref{Dnum})] for $w=2$~(a) and
  $w=4$~(b) ($L=1000$). The lines correspond to
  analytical calculations [Eq.~(\ref{Dmfin})]. The crosses ($\times$) with
  errorbars in~(a)
  are obtained from simulation data under quasi-equilibrium conditions [Eq.~(\ref{SS-D})].}\label{BI_st}
\end{figure*}
In order to obtain $D(\rho)$ from the steady-state data by using
Eq.~(\ref{Dnum}) we have measured the average current 
and the average density profile. The latter has been appropriately binned (the
density is averaged over $5$ sites), in order to be able to evaluate the derivative via
finite differences. The corresponding results are shown in
Figs.~\aref{Diff_staz},~\bref{Diff_staz}, and~\ref{BI_st}.
We note that $D(\rho \rightarrow 0) \rightarrow 1$. The correct behavior
at low densities is captured, while 
for large $\rho$ the data are still rather noisy, because the method to extract $D$ relies
on determining derivatives numerically.  
The overall agreement between results obtained from the spreading data and
from the steady-state data is
good for $\rho \gtrsim 0.5$ and even better for $\rho \gtrsim 2$, which shows that
the diffusion picture 
described in Subsec.~\ref{diff} leads to consistent results.

Simulations under quasi-equilibrium conditions have been restricted to the case $w=2$
because the quantitative agreement between the results obtained
from quasi-equilibrium, using Eq.~(\ref{SS-D}) for $D(\rho)$, and those
obtained in the steady state is satisfactory (see Fig.~\ref{BI_st}).    
Due to the small difference in density of the two reservoirs ($\delta
\rho = 0.01$) the average current is very small and requires accurate measurements.   
The data are obtained by averaging over $10^7$ configurations or
more (approximately $100$ times more than for the steady
state data), leading to a high precision for the density profile, too.
The autocorrelation time for the average density has been
carefully checked and the time intervals between samples have been chosen in order
to minimize correlations. The procedure allows one to estimate reliably the
statistical error for the diffusion coefficient, shown by the errorbars in
Fig.~\aref{BI_st}. Note that 
the results obtained from steady-state simulations with bigger reservoir
differences lie within the errorbars. 

\section{\label{CL} Continuum description}
In this section we derive the nonlinear diffusion equation
corresponding to the continuum limit of the model. The equation is derived
from the microscopic dynamics by using several simplifying assumptions.  
We start from the master equation, which describes exactly the dynamics of the
model and in its most general form can be written as
\begin{equation}\label{ME-gen}
\partial_t P_t(C) = \sum_{C'} \mathcal{M}(C,C') P_t(C'),
\end{equation}
where $C$ is a generic configuration, while $\mathcal{M}(C,C')$ encodes the transitions 
from the configuration $C$ to the configuration $C'$.
In our case, the operator $\mathcal{M}$ can be split into bulk ($\mathcal{M}_b$) and boundary
($\mathcal{M}_s$) terms, so that   
\begin{equation}\label{ME-2}
\partial_t P_t(C) =\sum_{C'} \left[ \mathcal{M}_s(C,C') +\mathcal{M}_b(C,C')\right] P_t(C').
\end{equation}
The operator $\mathcal{M}_b$ describes bulk moves, upon which particles are
exchanged between columns at sites $x \in [1,\dots,L]$ and which are associated with the rates
defined in Eq.~(\ref{rate3}).
The operator $\mathcal{M}_s$, describes boundary moves upon which particles
are inserted into
or removed from the system at the sites $x=1$ or $x=L$, and which are associated
with the rates introduced in Subsec.~\ref{BC}.
Explicit expressions for the operators $\mathcal{M}_b$ and $\mathcal{M}_s$ are given in Appendix~\ref{GKder}.

The evolution of the ensemble average of a generic (time-indipendent) operator $\mathcal{O}$ can be obtained from
Eq.~(\ref{ME-gen}) as
\begin{equation}
\partial_t \langle \mathcal{O} \rangle = \sum_{C,C'} \mathcal{O}(C) \mathcal{M}(C,C') P_t(C'),
\end{equation}
where $\mathcal{O}(C)$ is the value of the operator $\mathcal{O}$ for configuration $C$.
Recalling that $\mathcal{M}(C,C)=-\sum_{C' \neq C} \mathcal{M}(C',C)$, it is straightforward to obtain
\begin{equation}\label{genO-ev}
\partial_t \langle \mathcal{O} \rangle = \sum_{C} \mathcal{K}(C) P(C), 
\end{equation}
where $\mathcal{K}(C)$ is the jump moment of the operator $\mathcal{O}$ defined as
\begin{equation}\label{jumpm}
\mathcal{K}(C) \equiv \sum_{C' \neq C} \left[ \mathcal{O}(C')-\mathcal{O}(C) \right] \mathcal{M}(C',C).
\end{equation}
We consider now the operator $\mathcal{N}_x$, defined as $\mathcal{N}_x(C)=
n_x$. Accordingly, the configurations $C'$, 
for which the jump moments defined in Eq.~(\ref{jumpm}) are nonzero, are
$C'=\{n_1,\dots,n_{x-1}+1,n_x-1,n_{x+1},\dots,n_L\},
\{n_1,\dots,n_{x-1}-1,n_x,n_{x+1}+1,\dots,n_L\},
\{n_1,\dots,n_{x-1},n_{x}+1,n_{x+1}-1,\dots,n_L\},
\{n_1,\dots,n_{x-1},n_{x}-1,n_{x+1}+1,\dots,n_L\}$ with $x \in [2,\dots,L-1]$, so that
using Eqs.~(\ref{genO-ev}),~(\ref{jumpm}),~(\ref{Mact}), and~(\ref{Mb}) one
obtains in the bulk
\begin{equation}\label{MaE}
\begin{split}
& \partial_t \rho(x,t)  
= -\left[ \langle j(x+1,t) - j(x,t) \rangle \right], 
\end{split}
\end{equation}
where $\rho(x,t) = \langle n_x \rangle$, $x \in [2,\dots,L-1]$, and 
\begin{equation}\label{currates}
\langle j(x,t) \rangle  = \langle u(n_{x-1},n_x) - u(n_x,n_{x-1}) \rangle  
\end{equation}
is the mean local and instantaneous current in Eq.~(\ref{defcurr}).
Assuming that the probability distribution [Eq.~(\ref{ME-2})] 
factorizes completely, i.e., within the mean field approximation  
\begin{equation}
P(C,t) = \prod_{y=1}^{L} \tilde{p}_y(n_y,t),
\end{equation}
 the average rates in Eq.~(\ref{currates}) [for the definition of the
 rates see Eq.~(\ref{rate3})] reduce to
\begin{equation}\label{avrt}
\begin{split}
 & \langle  u\left( n_x, n_{x+1} \right) \rangle  \equiv \sum_{n_1=1}^{\infty} \dots
 \sum_{n_L=1}^{\infty} \\
 & \prod_{y=1}^{L} \tilde{p}_y(n_y,t) \left\{ n_x
 \, \exp{\left[ \frac{w}{(n_{x+1}+1)^4}-\frac{w}{n_x^4} \right]} \right\}  \\
 & = \tilde{f}_1(w,t,x) \tilde{f}_2(w,t,x+1), 
\end{split}
\end{equation}
where
\begin{subequations}
\begin{equation}
\tilde{f}_1(w,t,x) =  \sum_{n_x=1}^\infty \tilde{p}_x(n_x,t)  \, n_x
\exp{\left(\frac{w}{n_x^4}\right)}
\end{equation}
and
\begin{equation}
\tilde{f}_2(w,t,x) = \sum_{n_x=0}^\infty \tilde{p}_x(n_x,t) \exp{\left[-\frac{w}{(n_x+1)^4}\right]}.
\end{equation}
\end{subequations}
The second equation in Eq.~(\ref{avrt}) holds because
 the sum in the first line equals $1$ for any $y \neq x,x+1$  due to the
 normalization of the distribution $\tilde{p}$. 
 We further assume that the distribution $\tilde{p}_x$ depends smoothly on $x$ and
 $t$, and that it does so only via the mean site density $\rho(x,t)$ (or, equivalently, on an effective
local chemical potential) such that
$\tilde{p}_x(n,t) = p\left(n,\rho(x,t)\right)$ and $\tilde{f}_{1,2}(w,t,x) =
f_{1,2}\left(w,\rho(x,t)\right)$.  
Expanding the density up to second order in $a$ [i.e., $\rho(x+a) \approx \rho(x) + a \partial_x \rho +
\frac{1}{2} a^2 \, \partial^2_x \rho $] and then setting $a=1$
 [Eqs.~(\ref{MaE}) and~(\ref{avrt})], leads to the diffusion equation 
\begin{equation}
\begin{split}
\partial_t \rho(x,t) =  &\left[f_1(w,\rho) \partial_{\rho} f_2(w,\rho) - 
f_2(w,\rho) \partial_{\rho} f_1(w,\rho)\right] \\
& \times \partial_x^2  \rho(x,t),
\end{split}
\end{equation}
with a density-dependent diffusion coefficient
\begin{equation}\label{diff_d}
D(w,\rho) = 
f_1(w,\rho) \partial_{\rho} f_2(w,\rho) - f_2(w,\rho) \partial_{\rho} f_1(w,\rho).
\end{equation}
In the steady-state regime the functions $f_{1,2}$ can be computed explicitly
in the \emph{local equilibrium} approximation, i.e., by approximating the exact steady-state
distribution with a grand canonical \emph{equilibrium} Gibbs distribution $P_G$:
\begin{equation}\label{gw}
P_G(n,w,\mu_x) = \frac{1}{Z(w,\mu_x)}  \frac{1}{n!} 
\exp{\left(2 w \, h(n) - \mu_x n \right)}
\end{equation}
where
\begin{equation}\label{Z}
Z(w,\mu_x) = \sum_{n=0}^{\infty} \frac{1}{n!}
\exp{\left( 2 w h(n)  -\mu_x n \right)};
\end{equation}
$h$ is defined in Eq.~(\ref{HN}) and, as before,
$\mu_x=\beta \tilde{\mu}_x$ is a dimensionless chemical potential. These
approximations are expected to
hold if the density varies slowly in space, so that the
parts of the system to the left and to the right of $x$
act on the column at $x$ effectively as a particle reservoir with a
well-defined chemical potential $\mu_x$.   Accordingly, a nontrivial profile
$\rho_x$ emerges which smoothly interpolates between the reservoir
densities at $x=0$ and $x=L+1$ 

The distribution $P_G$ can be expressed in terms of the local density
$\rho_x$ by solving the implicit equation
\begin{equation}\label{nmueq}
\left< n_x(w) \right> = \sum_{n=1}^{\infty} P_G(n,w,\mu_x) \, n = \rho_x
\end{equation}
for $\mu_x = \mu(\rho_x)$.
The functions $f_1$ and $f_2$ in Eq.~(\ref{diff_d}) are then given by         
\begin{subequations}\label{2funct}
\begin{equation}\label{f1}
\begin{split}
& f_1(w,\rho) =  \frac{1}{Z(w,\mu)} \sum_{n \geq 0}  \\ 
& \hfill \left. \frac{1}{n!} \exp{ \left[2 w \, h(n) 
-\frac{w}{(n+1)^4} - \mu n \right]}\right|_{\mu=\mu(\rho)},
\end{split}
\end{equation}
\begin{equation}\label{f2}
\begin{split}
& f_2(w,\rho) =  \frac{1}{Z(w,\mu)} \sum_{n \geq 1} \\ 
& \hfill \left. \frac{1}{(n-1)!} \, 
\exp{\left(2 w \, h(n) + \frac{w}{n^4} -\mu n \right)}\right|_{\mu=\mu(\rho)}.
\end{split}
\end{equation}
\end{subequations}
Due to Eq.(\ref{HN}) one has
\begin{equation}
2w \, h(n) + \frac{w}{(n+1)^4} = 2w \, h(n+1) - \frac{w}{(n+1)^4},
\end{equation}
which implies
\begin{equation}
f_2 = \rme^{-\mu} f_1
\end{equation}
and thus
\begin{equation}\label{eq1}
\partial_\mu f_2  =  \rme^{-\mu} \partial_\mu f_1  - f_2.
\end{equation}
From Eq.~(\ref{nmueq}) one can infer the derivative of $\rho$ with respect to $\mu$:
\begin{equation}\label{eq2}
\begin{split}
\partial_\mu \rho & = \sum_{n=1}^{\infty} n 
\partial_\mu P_G = \sum_{n=1}^{\infty} n \,(\left< n  \right> - n)\, P_G  =  \\
&  = \left< n \right>^2 - \left< n^2 \right> \equiv -\chi(w,\mu).
\end{split}
\end{equation}
Combining Eqs.~(\ref{diff_d}),~(\ref{eq1}), and~(\ref{eq2}) one obtains 
\begin{equation}\label{Dmfin}
\begin{split}
D(w,\rho) &= \left. \frac{ f_1(w,\mu) f_2(w,\mu)}
{\chi(w,\mu)}\right|_{\mu=\mu(\rho)} =\\ 
& =\left. \frac{\langle u\left(
    n_x,n_{x+1} \right) \rangle }{\chi(w,\mu)}\right|_{\mu=\mu(\rho)}.
\end{split}
\end{equation}
Equation~(\ref{Dmfin}) allows one to compute numerically the diffusion
coefficient. To this end we solve Eq.~(\ref{nmueq}) for $\mu(\rho)$ and insert the solution 
$\mu(\rho(x))$ back into Eq.~(\ref{Dmfin}).
The resulting $D(\rho)$ is obtained by approximating the series in Eq.~(\ref{2funct}) by
finite sums. We have checked the stability of the calculation for 
densities  $0 < \rho < 11$, in order to be able to make contact with our Monte Carlo
data.  For $w < 1.5$ the theoretical expression is in good agreement with the
simulation data (see Fig.~\ref{Diff_staz}), but
for larger values of the  
interaction significant deviations occur (see Fig.~\ref{BI_st}). These
deviations  systematically increase with increasing interaction strength,
which cannot be easily blamed on numerical inaccuracies. A first possible explanation for the
deviations could be the non-equilibrium character of the simulations, due to the
chemical potential gradient present in the system.   
However, the quasi-equilibrium simulation results allow us to rule out strong
non-equilibrium effects as the primary source for these
deviations, because $D(\rho)$ computed from these data coincides with the one
obtained from non-equilibrium simulations; thus, the diffusion coefficient is
basically independent of the difference between the reservoirs densities. 
Taking advantage of this fact we can make use of general results for an
infinitely large system, derived in
quasi-equilibrium conditions, such as the Green-Kubo formula~\cite{Spohn-book}:       
\begin{equation}\label{GK}
\begin{split}
 D(\rho) = \frac{1}{\chi} & \Bigg[  \langle u(n_x,n_{x+1}) \rangle_{eq} \\ 
& \left. - \sum_{x'=1}^{\infty} \int_0^\infty \rmd t' \, \langle j_{x+1} U_{eq}(t') j_{x'}
   \rangle_{eq} \right],
\end{split}
\end{equation}   
where $\langle \cdot \rangle_{eq}$ indicates the average performed over the
equilibrium distribution and $U_{eq}$ is the evolution operator
for the equilibrium dynamics. Following the discussion
in Ref.~\cite{Spohn-book}, in Appendix~\ref{GKder} we present a brief
derivation of Eq.~(\ref{GK}) for the class of models we are interested in. 
The comparison of Eqs.~(\ref{Dmfin}) and~(\ref{GK}) shows that our mean-field
calculation reproduces the first term in Eq.~(\ref{GK}), while the terms which
would reduce (in the appropriate limit) to the time integral of the current correlations 
cannot be captured by this mean-field approximation. 

Computing explicitly the current correlations is difficult, but some
of their general properties~\cite{Spohn-book} allow us to conclude that they provide a
qualitatively correct correction to the diffusion coefficient calculated
via Eq.~(\ref{Dmfin}).     
 The function $\langle j_{x+1} U_{eq}(t)j_{x'} \rangle_{eq}$ appearing in Eq.~(\ref{GK}) is 
integrable, positive, and decaying exponentially for $t \rightarrow\infty$.
Thus it is a \emph{negative} 
contribution to Eq.~(\ref{GK}) and \emph{decreases} the diffusion coefficient 
obtained from Eq.~(\ref{Dmfin}). This is in qualitative agreement with the
data shown in Fig.~\ref{BI_st}.
 
\section{\label{Conc} Summary and conclusions}

We have introduced a lattice model (Fig.~\ref{Config}) for spreading of a fluid in narrow, quasi
  one-dimensional slit-like channels and in contact with particle reservoirs
  located at their ends. The model 
  accounts for long-range attractive substrate-fluid interactions, while the fluid-fluid
  interaction is taken to be hard-core only. We have studied the spreading
behavior and stationary state using kinetic Monte Carlo simulations and a
non-linear diffusion equation corresponding to the continuum limit of our
  discrete model. The main
results are the following. 

The spreading regime has been studied starting from an empty lattice:
  we have set the right reservoir density to $\rho_{L+1}=0$, so that it acts 
  as a perfect sink, and the left reservoir to a nonzero value (typically
  $\rho_0=11$), thereby feeding particles into the system. At intermediate
  times, for which the reservoirs do not play a relevant role, we have found a
  diffusion-like behavior, such that profiles at 
  different times collapse onto a single master curve if the scaling
  variable $\lambda=x / \sqrt{t}$ is introduced (see Fig.~\ref{Mv_Profs}). This
  scaling is further confirmed by the time evolution of the total mass [see
  Eq.~(\ref{M})] shown in Fig.~\ref{mass}.  In
  the steady state (Fig.~\ref{St_Profs}) we have found nontrivial density profiles
  depending on the substrate interaction strength. The dynamics at the boundary
  influences the density profiles: the analysis in Subsec.~\ref{BC}
  shows that one of the possible definitions of the boundary rates causes 
  discontinuities in the profiles. These discontinuities can be eliminated by an alternative
  definition or can be calculated explicitly (see Fig.~\ref{eq_dens}).  The
  profiles also depend on the reservoir density as discussed 
  in Appendix~\ref{App2}, 
  either lying completely above the free-diffusion straight line or
  crossing this line at a certain point. This feature can be described by the
  deviation $\Delta x(\rho)$ from free diffusion [see Eq.~(\ref{def_delta}) and
  Fig.~\ref{sketch}] and explained in terms of the function $R(\rho)$ introduced in
  Eq.~(\ref{acca}). In Fig.~\ref{ennesima} the general behavior of this
  function is sketched while in Fig.~\ref{res_infl} the simulation and mean-field
  results for the deviation $\Delta x(\rho)$ are shown.  

Assuming that a nonlinear diffusion equation describes the behavior of the
  particle density, we have extracted the diffusion coefficient
  from the simulation data in the spreading regime (Fig.~\aref{Diff_staz}) and
  in the steady state
  (Figs.~\ref{Diff_staz} and~\ref{BI_st}). The two sets of results
  are compatible with each other and show that the interaction with the
  substrate tends to enhance diffusion, as expected from qualitative
  arguments discussed in Subsec.~\ref{rates}. 

The Monte Carlo results for the diffusion coefficient are in agreement with an analytical calculation
  based on local equilibrium assumptions, in which the equilibrium grand canonical 
  distribution is modified by a spatially varying local chemical
  potential. The agreement is very good for weak interactions with the substrate
  (Fig.~\ref{Diff_staz}), but deteriorates for 
  strong interactions (Fig.~\ref{BI_st}). The independence of the diffusion coefficient from
  the  difference between the reservoirs densities, suggested by quasi-equilibrium
  simulations, allows one to qualitatively explain these deviations by using a
  Green-Kubo formula.

An experimental situation to which our model would apply should be effectively
quasi one-dimensional, with effective inert side walls 
confining the fluid and the substrate-fluid interaction strongly dominating over the
fluid-fluid interaction. This might be realized, e.g., by colloidal particles
in channels.    
An experimental investigation of the diffusion coefficient as a function of the 
density would allow one to make a direct comparison with the results presented here.     

\appendix

\section{\label{GKder} A Green-Kubo formula}
 
In this appendix we derive a Green-Kubo formula 
in the case of the non-equilibrium steady state induced by an infinitesimal
difference $\delta_\mu  =\mu_0
- \mu_{L+1} \rightarrow 0$ between the dimensionless (in units of $k_B T$)
chemical potentials $\mu_0$ and $\mu_{L+1}$ of the left and the right reservoir, respectively.  
The rates are denoted by $u(n_x,n_{x+1})$ but they are not necessarily of the explicit form
of Eq.~(\ref{rate3}). 
The master equation, which describes the dynamics of the system, is
given by Eq.~(\ref{ME-2}), where $\mathcal{M}_b$ is an operator that acts on a
distribution $P$ as
\begin{widetext}
\begin{equation}\label{Mact}
\begin{split}
&\sum_{C'} \mathcal{M}_b(C,C') P(C') =  \sum_{x=1}^{L-1} \left\{ -\left[ u\left(n_x,n_{x+1}\right)
+ u\left(n_{x+1},n_x \right) \right] P\left({n_1,\dots,n_L}\right) +  \right.\\ 
& \left. + u\left( n_{x+1}+1,n_x-1 \right)
P\left({n_1,\dots,n_x-1,n_{x+1}+1,\dots,n_L}\right) \right.\\
 &\left.+  u\left(n_x+1,n_{x+1}-1 \right)
 P\left({n_1,\dots,n_x+1,n_{x+1}-1,\dots,n_L}\right) \right\},
\end{split}
\end{equation}
\end{widetext}
and where $\mathcal{M}_s$ acts on $P$ as
\begin{widetext}
\begin{equation}\label{Mb}
\begin{split}
 \mathcal{M}_{s}(C,C') P(C') = -\left[ u\left(n_1,n_0 \right) + u\left(n_0,n_1 \right) +
  u\left(n_L,n_{L+1} \right) + u\left(n_{L+1},n_{L} \right)\right] 
P\left({n_1,\dots,n_L}\right) & \\
 \hfill + u\left( n_0,n_1-1 \right) P\left({n_1-1,\dots,n_L}\right) 
+ u\left(n_1+1,n_0 \right) P\left({n_1+1,\dots,n_L}\right) &\\
 \hfill + u\left( n_{L}+1,n_{L+1} \right) P\left({n_1,\dots,n_L+1}\right) 
+ u\left(n_{L+1},n_{L}-1 \right) P\left({n_1,\dots,n_L-1}\right) &.  
\end{split}
\end{equation}
\end{widetext}
If the alternative definitions of the rates at the boundaries
 are employed [Eq.~(\ref{Grates})] the rates $u_{\alpha,\kappa,\gamma,\delta}$ 
replace $u\left( n_0,n_1 \right), u\left( n_1,n_0 \right),$ and $u\left(
 n_L,n_{L+1} \right), u\left(n_{L+1},n_L \right) $ in Eq.~(\ref{Mb}).  

We require that the rates $u$ satisfy detailed balance, so that for
$\delta_\mu=0$, i.e., in the equilibrium case and thus without
net transport of particles, the system is described by the grand canonical 
distribution 
\begin{equation*}
P^{GC}_{eq} = P^C_{eq} \, \rme^{-\mu_0 N},
\end{equation*}
where $P^{C}_{eq}$ is a canonical distribution and
 $N=\sum_{x=1}^{L} n_x$ is the total number of particles in the system.
In the general case $\delta_\mu \neq 0$ detailed balance should be 
 satisfied \emph{locally} by the boundary rates at each end of the system.
 
The formal solution of the master equation~(\ref{ME-2}) is
\begin{equation}\label{ME-sol}
P(t) = \rme^{\mathcal{M} t} P_{in} = U(t) P_{in},
\end{equation}
where $P_{in}$ is the initial distribution $P(t=0)$ and $U(t)=\rme^{\mathcal{M} t}$ is
the time evolution operator. In the equilibrium case $\delta_\mu=0$  the corresponding
dynamics is indicated by $\mathcal{M}_{eq}$.  
Splitting arbitrarily the operator $\mathcal{M}$ as $\mathcal{M}=\mathcal{M}_0+\mathcal{M}_I$, one can check that
from the equation
\begin{equation}\label{Dyson}
U(t) = U_0(t) + \int_0^t \rmd t' U(t-t') \mathcal{M}_I  U_0(t'), 
\end{equation}
where $U_0(t) = \rme^{\mathcal{M}_0 t}$, the two equations
\begin{equation}\label{boh}
\begin{split}
& \partial_t U(t) = \mathcal{M} U(t) \\
& U(0) = 1
\end{split}
\end{equation}
can be obtained. Since these equations have as unique solution the evolution
operator $U(t)=\rme^{\mathcal{M} t}$, Eq.~(\ref{Dyson}) holds.
Assuming that unique stationary distributions $P_0$ and $P_{st}$ exist for
the dynamics defined by $\mathcal{M}_0$ and $\mathcal{M}$, respectively, i.e.,
\begin{equation}\label{eq-steady}
\begin{split}
& \mathcal{M}_0 P_0 = 0, \\
& \mathcal{M} P_{st} = 0,
\end{split} 
\end{equation}
Eqs.~(\ref{Dyson}) and~(\ref{eq-steady}) lead to
\begin{equation}\label{Dys-dis}
P_{st} = \lim_{t \rightarrow \infty } \left[ U(t) P_0 \right] =P_0 + \int_0^\infty \rmd t \, U(t) \mathcal{M} P_0.
\end{equation}
The arbitrariness in the splitting of $\mathcal{M}$ allows one to choose $P_0$ as the local
equilibrium distribution
\begin{equation}\label{l-eq}
P_0 =  P_{eq}^{C} \exp{\left[-\sum_{x=1}^{L} \mu(x) n(x) \right]}.
\end{equation}
For $\delta_\mu \ll 1$ one has  $\mu(x) \simeq \mu + \delta_\mu (x/L)$; applying 
 $\mathcal{M}$ to $P_0$ [see Eq.~(\ref{Mact})] and by using the detailed balance condition one obtains
\begin{equation}
\begin{split}
\mathcal{M} P_0 = \sum_{x=1}^{L-1} & \left\{ u(n_x,n_{x+1})  \left[
    \rme^{-\frac{\delta_\mu}{L}}-1 \right]  \right. \\
& \left.  +u(n_{x+1},n_x) \left[ \rme^{\frac{\delta_\mu}{L}}-1 \right] \right\}
P_0(\{n_1,n_{L}\}).
\end{split}
\end{equation}
The aim is to calculate the mean current $\langle j \rangle_{st}$, from which one can
obtain $D$ in the limit of small density differences by using
Eq.~(\ref{SS-D}). Averaging $j$, given by Eq.~(\ref{currates}), over the
distribution $P_{st}$ and expanding to first order in $\delta$ leads to 
\begin{equation}
\begin{split}
\langle j_{x+1} \rangle_{st} =  \frac{\delta_\mu}{L} & \Bigg[ \langle
  u(n_x,n_{x+1}) \rangle_{0} \,  \\
& \left. - \sum_{x'=1}^{L} \int_0^\infty \rmd t'\, \langle j_{x+1} U(t') j_{x'} \rangle_0
\right] + O\left(\frac{\delta_\mu^2}{L^2} \right).
\end{split}
\end{equation}
where $\langle \cdot \rangle_0$ indicate the average taken over the local equilibrium
distribution [Eq.~(\ref{l-eq})].
By taking the limits $\delta_\mu \rightarrow 0$ and $L\rightarrow \infty$ we
obtain the Green-Kubo formula
\begin{equation}\label{eq:GreenKubo}
\begin{split}
 D(\rho) = \frac{1}{\chi} & \Bigg[  \langle u(n_x,n_{x+1}) \rangle_{eq} \\ 
 & \left.- \sum_{x'=1}^{\infty} \int_0^\infty \rmd t' \, \langle j_{x+1} U_{eq}(t') j_{x'}
   \rangle_{eq} \right]
\end{split}
\end{equation}
with $\chi=\langle n_x^2 \rangle_{eq}-\langle n_x \rangle_{eq}^2$.
The limits $\delta_\mu \rightarrow 0$ and $L\rightarrow \infty$ imply that $P_0 \rightarrow P_{eq}$ and $U_0
\rightarrow U_{eq}=\rme^{\mathcal{M}_{eq} t}$.
Note that in Eq.~(\ref{eq:GreenKubo}) the averages are taken over the equilibrium
distribution and the properties of the diffusion  
coefficient are completely determined by the \emph{equilibrium} dynamics of
the model. Accordingly, the r.h.s. of Eq.~(\ref{eq:GreenKubo}) is independent
of $x$ and thus $D$ depends on $\rho$ only.

\section{\label{App2} Influence of the reservoir densities on the density profiles}

\begin{figure}[hc]
\centering
\epsfig{figure=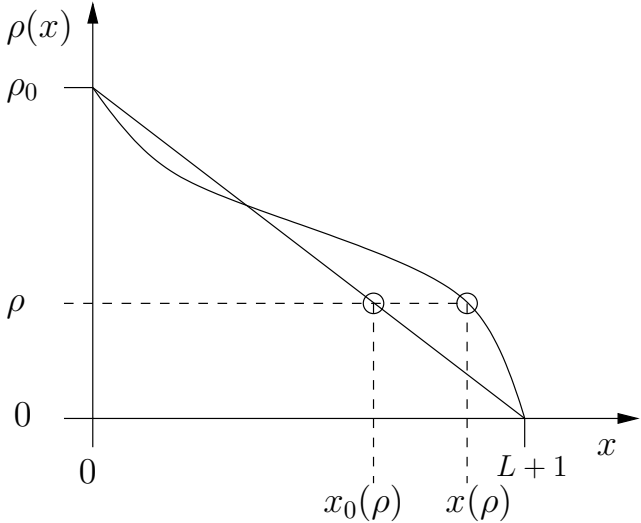,width=\columnwidth}
\caption{Schematic comparison between a typical steady-state density profile (curved line) 
 and free diffusion (straight line). At a given density $\rho$ the
 corresponding positions $x(\rho)$ and $x_0(\rho)$ are
 taken from the steady-state density profile and from the free-diffusion
 line, respectively, and $\Delta x(\rho)$ is determined from Eq.~(\ref{def_delta}).
 }\label{sketch}. 
\end{figure}

In Sec.~\ref{furt_an} we have shown that, within our model, the density-dependent diffusion
coefficient $D(\rho)$ does not depend on the boundary conditions,
i.e., the densities of the left and right reservoirs. 
As shown in Fig.~\ref{St_Profs}, the density profiles for low 
reservoir densities seemingly lie mostly above the free diffusion straight line [see
Fig.~\bref{St_Profs}], whereas the opposite happens for high reservoir densities [see
Fig.~\aref{St_Profs}]. While these profiles refer to  different interaction
strength $w$, we note that the difference between $D(\rho)$ for
$w=2$ and for $w=1.5$ [see Figs.~\bref{Diff_staz} and~\aref{BI_st}] is not
very marked. Therefore it is unlikely that this difference explains the
differences between 
the profile corresponding to $w=1.5$ in Fig.~\aref{St_Profs} and the profile
corresponding to $w=2$ in Fig.~\bref{St_Profs}.

In order to clarify the relation between the steady-state profile and the
boundary conditions, here we restrict our analysis to 
the case in which the right reservoir is a perfect sink,
($\rho_{L+1}=0$) and the left reservoir density $\rho_0$ is varied, keeping
the interaction strength $w$ constant. 
The definitions in Eq.~(\ref{Grates}) for the rates at the boundaries have been
used, so that the density profile is a continuous function of $x$.

\begin{figure}[hc]
\centering
\epsfig{figure=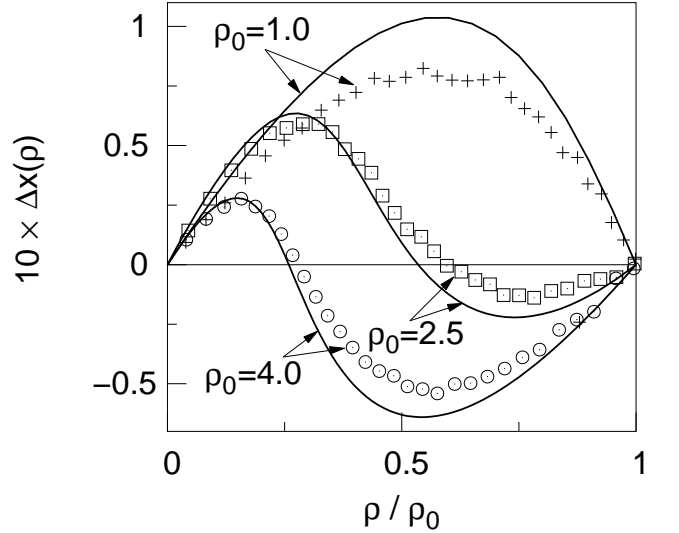,width=\columnwidth}
\caption{$\Delta x(\rho)$ [see Eq.~(\ref{def_delta})] for left reservoir densities $\rho_0=1$ ($+$), $\rho_0=2.5$
 ($\boxdot$) and $\rho_0=4$ ($\odot$) with fixed $w=1.5$, $L=300$,
  and $\rho_{L+1}=0$.
 The full lines correspond to  $\Delta x(\rho)$ obtained by using
 the mean-field diffusion coefficient [Eq.~(\ref{Dmfin}]. The quantity
 $\Delta x(\rho)$ in the $y$-axis is multiplied by a 
 factor $10$ in order to enhance visibility.}\label{res_infl}. 
\end{figure}

A convenient quantity to describe the dependence on $\rho_0$ of the profiles
is the deviation in $x$ at fixed 
$\rho$ between the actual density profile and the straight line corresponding to
free diffusion (see Fig.~\ref{sketch}), defined as
\begin{equation}\label{def_delta}
 \Delta x (\rho)= \frac{x(\rho) - x_0(\rho)}{L+1},
\end{equation}
where $x(\rho)$ is the inverted steady-state profile and $x_0(\rho)=(L+1)(1-\rho/\rho_0)$ is the
corresponding free diffusion line. In Fig.~\ref{res_infl} we report the
numerical data for $\Delta x (\rho)$ corresponding to three
left reservoir densities ($\rho_0=1$, $2.5$, and $4$). These results exhibit the typical behavior of
the deviation $\Delta x$: at small $\rho_0$ it is positive for all $\rho$ ($\rho_0=1$)
whereas upon increasing the reservoir density a negative part appears
($\rho_0=2.5$ and $4$). 
This pattern can be explained along the line of arguments introduced in
Sec.~\ref{furt_an} based on the coarse-grained description of the
diffusion-like behavior. The nonlinear diffusion equation [Eq.~(\ref{NL-diff})]
yields for the inverse steady-state profile $x(\rho)$ 
\begin{equation}\label{inv_diff}
\frac{\rmd x(\rho)}{\rmd \rho}   = - \frac{D(\rho)}{j},
\end{equation}
where $j$ is the current and $D$ the diffusion coefficient. Integrating
Eq.~(\ref{inv_diff}) one obtains 
\begin{equation}\label{xsol_1}
x(\rho)   = \frac{1}{j} \int_\rho^{\rho_0}\rmd \xi D(\xi),
\end{equation}
where $0 \leq \rho \leq \rho_0$. Imposing the boundary condition $x(0)=L+1$ at the right
end of the system leads to  
\begin{equation}\label{jsol}
j   = \frac{1}{L+1} \int_0^{\rho_0}\rmd \xi D(\xi),
\end{equation}
so that $x(\rho)$ is given by
\begin{equation}\label{xsol_2}
x(\rho)   = (L+1) \left[\int_0^{\rho_0} \rmd \xi D(\xi) \right]^{-1}
\int_\rho^{\rho_0} \rmd \xi D(\xi).
\end{equation}
By combining Eqs.~(\ref{def_delta}) and~(\ref{xsol_2}) one obtains 
\begin{equation}\label{Deltax_1}
\begin{split}
 \Delta x (\rho) &= \frac{x(\rho) - x_0(\rho)}{L+1}  =\\ 
&=\left[\int_0^{\rho_0} \rmd \xi D(\xi)  \right]^{-1}
\int_\rho^{\rho_0} \rmd \xi D(\xi)  - \frac{\rho_0-\rho}{\rho_0},
\end{split}
\end{equation}
which, as expected, vanishes for $\rho=0$ and $\rho=\rho_0$. One can
re-write the last equation as follows:
 \begin{equation}\label{Deltax_2}
\begin{split}
 \Delta x (\rho) =  
\rho & \left[\int_0^{\rho_0} D(\xi) \rmd \xi \right]^{-1} \\ 
 &\times \left[ \frac{1}{\rho_0} \int_0^{\rho_0} \rmd \xi D(\xi) - \frac{1}{\rho} \int_0^{\rho} \rmd \xi D(\xi) 
 \right]. 
\end{split}
\end{equation}
Note that $\rho \left[\int_0^{\rho_0} D(\xi) \rmd \xi \right]^{-1}$ is
nonnegative whereas the second factor on  the r.h.s. of
Eq.~(\ref{Deltax_2}) can change 
sign. Denoting this latter factor by $h(\rho)$ and introducing the function
$f(\rho) = D(\rho) -1$, one obtains
  \begin{equation}\label{acca}
 h(\rho) = 
\frac{1}{\rho_0} \int_0^{\rho_0} \rmd \xi f(\xi) - \frac{1}{\rho}
\int_0^{\rho} \rmd \xi f(\xi) = R(\rho_0) - R(\rho),
\end{equation}
where $R(\rho)=\frac{1}{\rho} \int_0^{\rho} \rmd \xi \left[D(\xi)-1 \right]$.
Within our model the function $f$ is always positive (apart from Fig.~\aref{Diff_staz} for
$\rho \rightarrow 0$ and some noisy data at large $\rho$ in
Figs.~\ref{Diff_staz} and~\ref{BI_st}) and it vanishes for $\rho 
\rightarrow 0$ and $\rho \rightarrow \infty$. Thus $R(\rho)$ also vanishes 
for $\rho \rightarrow 0$ or $ \rho \rightarrow \infty$ and hence has a maximum
at a certain $\rho_{M}$, with $0<\rho_{M}<\infty$.  
\begin{figure}
\centering
\epsfig{figure=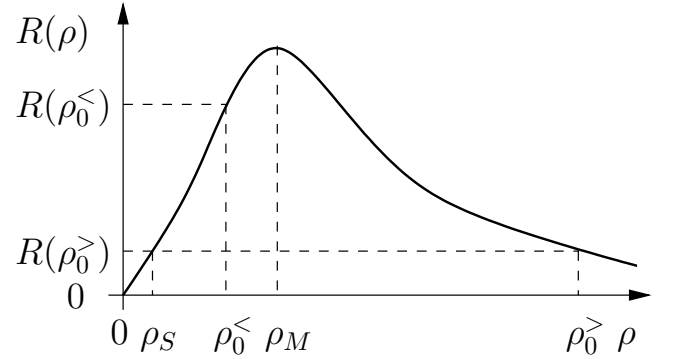,width=\columnwidth}
\caption{Qualitative plot of the function $R(\rho)$ (see Eq.~(\ref{acca})). Two possible
  situations are shown: if the reservoir density $\rho_0^<$ is smaller than
  $\rho_M$, there are no densities $ 0 \leq \rho < \rho_0^<$ such that
  $R(\rho)=R(\rho_0^<)$. In the opposite case with reservoir density $\rho_0^>
  > \rho_M$, there exist a density $0 < \rho_S < \rho_0^>$ such that
  $R(\rho_S)=R(\rho_0^>)$. }\label{ennesima}. 
\end{figure} 
A sketch of this function is provided in Fig.~\ref{ennesima}. Two
qualitatively different behaviors emerge by varying the left reservoir density $\rho_0$.
If the reservoir
density is set to  $\rho_0^< \leq \rho_M$, then in the
interval $0 \leq \rho \leq \rho_0^<$  no solution to $R(\rho_0^<) -
R(\rho)=0$ can be found other than the trivial one $\rho = \rho_0^<$. 
In this situation, we have
$R(\rho) < R(\rho_0^<)$ so that the function $\Delta x(\rho)$ is always
positive in the considered interval of densities, and the density
profile always lies above the line corresponding to free diffusion. This is
the case for 
$\rho_0=1$ and substrate interaction $w=1.5$ as shown in Fig.~\ref{res_infl}.   
In the opposite situation, in which the density of the
reservoir is set to $\rho_0^> \geq \rho_M$, there always exists a density
$\rho_S < \rho_0^>$ such that $R(\rho_0^>) - R(\rho_S)=0$. At this density
the density profile crosses the free diffusion line. For $\rho >
\rho_S$, $\Delta x(\rho)>0$ while $\Delta x(\rho)<0$ for $\rho <
\rho_S$. In Fig.~\ref{res_infl} this corresponds to the cases $\rho_0=2.5$ and
$\rho_0=4$.


\begin{thebibliography}{05}
\bibitem{Giord_Cheng}  N.~Giordano and J.T.~Cheng, J. Phys: Condens. Matter
  {\bf 13}, R271  (2001).
\bibitem{SeCaMa} P.R.~Selvaganapathy, E.T.~Carlen, and C.H.~Mastrangelo,
  Proceedings of the IEEE {\bf 91}, 954  (2003).
\bibitem{Mukho} R.~Mukhopadhyay, Analy. Chem. {\bf 78}, 7379 (2006). 
\bibitem{Pears_Cumm} J.L.~Pearson and D.R.S.~Cumming,
 Micro. Eng.  {\bf 78-79}, 343 (2005).
\bibitem{Rev_Fabr}  D.~Mijatovich, J.C.T.~Eijkel, and A.~van~den~Berg, Lab on a
  Chip {\bf 5}, 492 (2004).
\bibitem{Supple_Quirke_03} S.~Supple and N.~Quirke, Phys. Rev. Lett. {\bf 90},
  214501 (2003).
\bibitem{Hum_Ras_Now_01} G.~Hummer, J.C. Rasaiah, and J.P. Noworyta, Nature 
{\bf 414}, 188 (2001).
\bibitem{Dietr_Pop_Rau_05}  S.~Dietrich, M.N.~Popescu, and M.~Rauscher, J. Phys:
 Condens. Matter {\bf 17}, S577 (2005).   
\bibitem{MD_early} J.A.~Nieminen, D.B.~Abraham, M.~Karttunen, and K.~Kaski, 
 Phys. Rev. Lett. {\bf 69}, 124 (1992).
\bibitem{Gleb_PRE} G.~Oshanin, J.~De~Coninck, A.M.~Cazabat, and M.~Moreau,
 Phys. Rev. E {\bf 58}, R20 (1998).
\bibitem{V_DCon} M.~Vou\`e and J.~De Coninck, Acta Mater. {\bf 48}, 4405 (2000).
\bibitem{Kop_Lo_Dietr_Rausc} J.~Koplik, T.S.~Lo, M.~Rauscher, and S.~Dietrich, 
Phys. Fluids {\bf 3}, 032104  (2006).
\bibitem{Exp1}  F.~Heslot, A.M.~Cazabat, and N.~Fraysse,
  J. Phys.: Condens. Matter {\bf 1}, 5794 (1989). 
\bibitem{Exp2}  F.~Heslot, A.M.~Cazabat, and P.~Levinson,
Phys. Rev. Lett. {\bf 62}, 1286 (1989).
\bibitem{Exp3}   F.~Heslot, A.M. Cazabat, P.~Levinson, and N.~Fraysse,
Phys. Rev. Lett. {\bf 65}, 599 (1990).
\bibitem{Exp4}   J.~Daillant, J.J.~Benattar, and L.~Leger,
Phys. Rev. A {\bf 41}, 1963 (1990).
\bibitem{Exp5}   U.~Albrecht, A.~Otto, and P.~Leiderer,
Phys. Rev. Lett. {\bf 68}, 3192 (1992).
\bibitem{Exp6}   N.~Fraysse, M.P.~Valignant, F.~Heslot, A.~M.~Cazabat, and 
P.~Levinson, J. Colloid Interface Sci. {\bf 158}, 27 (1993).
\bibitem{Exp7}   M. Vou\`e, M.P.~Valignant, G.~Oshanin, A.~M.~Cazabat, and 
J.~De~Coninck, Langmuir {\bf 14}, 5951 (1998).
\bibitem{Exp8}   E.~P\'erez, E.~Sch\"affer, and U.~Steiner, 
J. Colloid Interface Sci {\bf 234}, 178 (2001).
\bibitem{Burlatsky} S.~F.~Burlatsky, G.~Oshanin, A.~M.~Cazabat, and M.~Moreau,
Phys. Rev. Lett.  {\bf 76}, 86 (1996).
\bibitem{Gleb_eal}  G.~Oshanin, J.~De~Coninck, A.M.~Cazabat, and M.~Moreau, 
J. Mol. Liq. {\bf 76}, 195 (1998).
\bibitem{Pop_Dietr_04} M.N.~Popescu and S.~Dietrich, Phys. Rev. E {\bf 69}, 
061602 (2004).
\bibitem{Pop_Dietr_Osh_05}  M.N.~Popescu, S.~Dietrich, and G.~Oshanin, 
J. Phys: Condens. Matter {\bf 17}, S4189 (2005).\bibitem{Chou1} T.~Chou, Phys. Rev. Lett. {\bf 80}, 85 (1998).  
\bibitem{Chou2} T.~Chou, J. Chem. Phys. {\bf 110}, 606 (1999).
\bibitem{Chou3} T.~Chou, Biophys. J. {\bf 86}, 2827 (2004).
\bibitem{Chou4} T.~Chou, J. Phys.: Math. Gen. A {\bf 39}, 2253 (2006).
\bibitem{MonPer} K.~K.~Mon and J.~K.~Percus, J. Chem. Phys. {\bf 119}, 3343
  (2003).
\bibitem{BhaJeNich} S.~K.~Bhatia, O.~Jepps, and D.~Nicholson, J. Chem. Phys. {\bf 120}, 4472
  (2004).
\bibitem{DemStSu} P.~Demontis, G.~Stara, and G.~Suffritti, J. Chem. Phys. {\bf 120}, 9233
  (2004).
\bibitem{MarTe} F.~Marchesoni and A.~Taloni, Phys. Rev. Lett. {\bf 97}, 106101
  (2006).
\bibitem{Bech} C.~Lutz, M.~Kollmann, P.~Leiderer, and C.~Bechinger, J. Phys.:
  Condens. Matter {\bf 16}, S4075 (2004).
\bibitem{troppiA} S.~M.~Saparov, J.~R.~Pfeifer, L.~Al-Momani, G.~Portella,
  B.~L.~de~Groot, U.~Koert, and P.~Pohl, Phys. Rev. Lett. {\bf 96}, 148101
  (2006).
\bibitem{Coc-thiv1} E.~Andjel, C.~Cocozza-Thivent, and M.~Roussignol,
 Ann. de l'I.Henry Poincar\'e {\bf 21}, 363 (1985).
\bibitem{Coc-thiv2} E.~Andjel, C.~Cocozza-Thivent, and M.~Roussignol,
 Z. f. Wahrscheinlichkeitstheorie {\bf 70}, 509 (1985).
\bibitem{Godr1} C.~Godr\`eche, preprint {\it cond-mat/0604276} (2006).
\bibitem{Godr2} C.~Godr\`eche and J.~M.~Luck, Eur. Phys. J. B {\bf 21}, 473 (2001).
\bibitem{Binder} K.~Binder, in \emph{Monte Carlo Methods in Statistical
    Physics}, edited by K.~Binder (Springer, Berlin, 1986), p. 30.
\bibitem{N-fold} A.B.~Bortz, M.H.~Kalos, and J.L.~Lebowitz, J. Comput.
  Phys. {\bf 17}, 10 (1975).
\bibitem{Jona} L.~Bertini, A.~da~Sole, D.~Gabrielli, G.~Jona-Lasinio, and
  C.~Landim, J. Stat. Phys. {\bf 107}, 635 (2002).
\bibitem{Spohn-book} H.~Spohn, \emph{Large scale dynamics of interacting
  particles} (Springer, Berlin, 1991).
\end{thebibliography}
\end{document}